\documentclass[a4paper,11pt]{amsart}

\usepackage[english]{babel}
\usepackage[T1]{fontenc}
\usepackage{lmodern}

\usepackage[utf8]{inputenc}
\usepackage{amsmath,amssymb,mathrsfs}
\usepackage{amscd}
\usepackage{pgf,tikz}
\usepackage{hyperref}
\usepackage{a4wide}

\numberwithin{equation}{section}
\newtheorem{theo}{Theorem}[section]
\newtheorem{cor}{Corollary}[section]

\newtheorem{prop}{Proposition}[section]
\newtheorem{lem}{Lemma}[section]
\newtheorem{de}{Definition}[section]
\newtheorem{rem}{Remark}[section]
\numberwithin{figure}{section}

\date{}

\newcommand{\ch}{{-\Delta}}

\newcommand{\cho}{{-\Delta}}
\newcommand{\bn}{{\mathbb{N}}}
\newcommand{\bz}{{\mathbb{Z}}}
\newcommand{\br}{{\mathbb{R}}}
\newcommand{\bc}{{\mathbb{C}}}
\newcommand{\bt}{{\mathbb{T}}}

\newcommand{\sd}{{\textbf{\textup{S}}_2}}
\newcommand{\sinf}{{\textbf{\textup{S}}_\infty}}

\newcommand{\im}{\operatorname{Im}}
\newcommand{\re}{\operatorname{Re}}

\newenvironment{prof}
	{\textbf{Proof.}}
	{\hfill $\blacksquare$\vskip 8pt}

\title{Resonances on regular tree graphs}

\author{Olivier \textsc{Bourget}}
\address{Departamento de Matem\'aticas, Facultad de Matem\'aticas, 
Pontificia Universidad Cat\'olica de Chile, Vicu\~na Mackenna 
4860, Santiago de Chile}

\email{obourget@mat.uc.cl}

\author{Diomba \textsc{Sambou}}
\address{Departamento de Matem\'aticas, Facultad de Matem\'aticas, 
Pontificia Universidad Cat\'olica de Chile, Vicu\~na Mackenna 
4860, Santiago de Chile}

\email{disambou@mat.uc.cl}

\author{Amal \textsc{Taarabt}}
\address{Departamento de Matem\'aticas, Facultad de Matem\'aticas, 
Pontificia Universidad Cat\'olica de Chile, Vicu\~na Mackenna 
4860, Santiago de Chile}

\email{amtaarabt@mat.uc.cl}


\begin{document}

\begin{abstract}
We investigate the distribution of the resonances near spectral thresholds
of Laplace operators on regular tree graphs with $k$-fold branching, $k \geq 1$,
perturbed by nonself-adjoint exponentially decaying potentials.
We establish results on the absence of resonances which in particular involve absence
of discrete spectrum near some sectors of the essential spectrum of the operators. 
\end{abstract}

\maketitle


\section{Introduction}\label{s1}


A great interest has been focused in the last decades on spectral analysis of Laplace 
operators on regular trees. This includes local perturbations \cite{al,af}, random settings \cite{kl,aw,ai,fr,fr1,sh} (see also the references therein), and quantum ergodicity regimes 
\cite{am}. For complementary results, we refer the reader to the papers \cite{br,bk,ro1,ro2,rr}, 
and for the relationships between the Laplace operator on trees and quantum graphs, see \cite{km}. 
However, it seems that resonances have not been systematically studied in the context of 
(regular) trees.

\smallskip

In this paper, we use resonance methods to obtain better understanding of local spectral 
properties for perturbed Schrödinger operators on regular tree graphs with $k$-fold branching, 
$k \geq 1$, as we describe below (cf. Section \ref{secd}). Our techniques are similar to 
those used in \cite{bbr1,bbr2} (and references therein), where self-adjoint perturbations
are considered. Actually, these methods can be extended to nonself-adjoint models, see for 
instance \cite{sa}. Here, we are focused on some nonself-adjoint perturbations of the Laplace 
operator on regular tree graphs. In particular, we shall derive as a by-product, a description 
of the eigenvalues distribution near the spectral thresholds of the operator. 

\smallskip

Since a nonself-adjoint framework is involved in this article, it is convenient to clarify the 
different notions of spectra we use. Let $T$ be a closed linear operator acting on a separable 
Hilbert space $\mathscr{H}$, and $z$ be an isolated point of $\sigma(T)$ the spectrum of $T$. 
If $\gamma$ is a small contour positively oriented containing $z$ as the only point of $\sigma(T)$, 
we recall that the Riesz projection $P_z$ associated to $z$ is defined by 
\begin{equation}\label{eq1,8}
P_z := \frac{1}{2i\pi} \oint_{\gamma} (T - \zeta)^{-1} d\zeta.
\end{equation}
The algebraic multiplicity of $z$ is then defined by 
\begin{equation}\label{eq1,9}
\textup{m}(z) := \text{rank} (P_z),
\end{equation}
and when it is finite, the point $z$ is called a discrete eigenvalue of the operator $T$. 
Note that we have the inequality $\mathrm{m} \, (z) \geq \mathrm{dim} \, \big( \text{Ker} 
(T - z) \big)$, which is the geometric multiplicity of $z$. The equality holds  
if $T = T^\ast$. So, we define the discrete spectrum of $T$ as
\begin{equation}\label{eq1,10}
\sigma_{\text{\bf disc}}(T) := \big\lbrace z \in \sigma(T) : z \hspace*{0.1cm} 
\textup{is a discrete eigenvalue of $T$} \big\rbrace.
\end{equation}
We recall that if a closed linear operator has a closed range and both its kernel and 
cokernel are finite-dimensional, then it is called a Fredholm operator. Hence, we define 
the essential spectrum of $T$ as
\begin{equation}\label{eq1,11}
\sigma_{\text{\bf ess}}(T) := \big\lbrace z \in \bc : \textup{$T - z$ \textup{is not a 
Fredholm operator}} \big\rbrace.
\end{equation} 
Note that $\sigma_{\text{\bf ess}}(T)$ is a closed subset of $\sigma(T)$.

\medskip

The paper is organized as follows. In Section \ref{secd}, we present our model.
In section \ref{secp}, we state our main results Theorem \ref{t1} and Corollaries
\ref{t2}, \ref{t3}. Section \ref{sec3} is devoted to preliminary results we need due 
to Allard and Froese. In Section \ref{sec4}, we establish a formula giving a kernel 
representation of the resolvent associated to the operator we consider and which is 
crucial for our analysis. In Section \ref{sec5}, we define and characterize 
the resonances near the spectral thresholds, while in Section \ref{sec6} we give the 
proof of our main results. Section \ref{sa} gathers useful tools on the characteristic 
values concept of finite meromorphic operator-valued functions.

\section{Presentation of the model}\label{secd}

We consider an infinite graph 
$\mathbb{G} = (\mathcal{V},\mathcal{E})$ with vertices $\mathcal{V}$ and edges $\mathcal{E}$, 
and we let $\ell^2(\mathcal{V})$ be the Hilbert space
\begin{equation}\label{eq1,0}
\ell^2(\mathcal{V}) := \bigg\lbrace \phi : \mathcal{V} \rightarrow \bc : 
\Vert \phi \Vert^2 := \sum_{v \in \mathcal{V}} \big\vert \phi(v) 
\big\vert^2 < \infty \bigg\rbrace,
\end{equation}
with the inner product 
\begin{equation}\label{eq,ps}
\langle \phi,\psi \rangle := \sum_{v \in \mathcal{V}} \phi(v) \overline{\psi}(v).
\end{equation}
On $\ell^2(\mathcal{V})$, we consider the symmetric Schrödinger operator $\cho$
defined by
\begin{equation}\label{eq1,1}
(\cho \phi)(v) := - \sum_{w \, : \, w \sim v} \Big( \phi(w) - \phi(v) \Big),
\end{equation}
where $w \sim v$ means that the vertices $w$ and $v$ are connected by an edge. If 
we define on $\ell^2(\mathcal{V})$ the symmetric operator $L$ by
\begin{equation}\label{eq1,2}
(L \phi)(v) := \sum_{w \, : \, w \sim v} \phi(w),
\end{equation}
then it is not difficult to see that the operator $\cho$ can be written as
\begin{equation}\label{eq1,3}
\cho = -L + d,
\end{equation}
where $d$ is the multiplication operator by the function (also) noted 
$d : \mathcal{V} \rightarrow \bc$, with $d(v)$ denoting the number of edges connected with
the vertex $v$. Note that when $d$ is bounded, then so is the symmetric operators $\cho$ 
and $L$, hence self-adjoint. In a regular rooted tree graph with $k$-fold branching, $k \ge 1$, 
(see Figure 2.1 for a binary tree graph), we have $d = k + 1 - d_0$ with
\begin{equation}
d_0(v) = \begin{cases} 
1 & \text{if } v \text{ coincides with the root of the tree}, \\ 
0 & \text{otherwise}. 
\end{cases}
\end{equation}
This is the same model described in \cite{af} and we refer to 
this paper for more details. 
In \eqref{eq1,3}, $d$ can be viewed as a perturbation of the operator $L$. It is well know 
(see Lemma \ref{l3,2}) that the spectrum of the operator $L$ is absolutely continuous, 
coincides with the essential spectrum and is equal to
\begin{equation}\label{eq1,4}
\sigma (L) = \sigma_{\textup{ac}} (L) = \sigma_{\textup{ess}} (L) = 
\left[ -2\sqrt{k},2\sqrt{k} \right].
\end{equation}



On $\ell^2(\mathcal{V})$, we define the perturbed operator
\begin{equation}\label{eq1,5}
\ch_M := \cho + M, 
\end{equation}
where $M$ is identified with the multiplication operator by the bounded potential function 
(also) noted $M$. In a regular rooted tree graph with $k$-fold branching, according to 
above, the operator $\ch_M$ can be written as
\begin{equation}\label{eq1,6}
\ch_M = -L + k + 1 - d_0 + M. 
\end{equation}
In \eqref{eq1,6}, the degree term $d_0$ can be included in the potential perturbation so 
that $\ch_M$ can be viewed as a perturbation of the operator $-L + k + 1$. Hence, from 
now on, the operator $\ch_M$ will be written as
\begin{equation}\label{eq1,60}
\ch_{\widetilde M} = -L + k + 1 + \widetilde M \quad {\rm with} \quad 
\widetilde M := - d_0 + M. 
\end{equation}
In the sequel, we set 
\begin{equation}\label{eq1,61}
t_\pm(k) := \pm 2\sqrt{k} + k + 1,
\end{equation}
and we shall simply write $t_\pm$ when no confusion can arise. Then, from \eqref{eq1,4}, 
it follows that the spectrum of the operator $-L + k + 1$ satisfies
\begin{equation}\label{eq1,7}
\sigma (-L + k + 1) = \sigma_{\textup{ac}} (-L + k + 1) = \sigma_{\textup{ess}} (-L + k + 1) = 
\left[ t_-(k),t_+(k) \right],
\end{equation}
where the $t_\pm(k)$ play the role of thresholds of this spectrum.

\smallskip

Now, let us choose some vertex $v_0 = 0 \in \mathcal{V}$ as the origin of the graph 
$\mathbb{G} = (\mathcal{V},\mathcal{E})$. For $v \in \mathcal{V}$, we define 
$\vert v \vert$ as the length of the shortest path connecting $0$ to $v$. Hence, 
$\vert v \vert$ defines in the graph the distance from $0$ to $v$. For $r > 0$, 
let $S_r$ be the sphere of radius $r$ in the graph defined by
\begin{equation}
S_r := \big\lbrace v \in \mathcal{V} : \vert v \vert = r \big\rbrace.
\end{equation}


\begin{figure}[h]\label{fig 1}
\begin{center}

\includegraphics[scale=0.8]{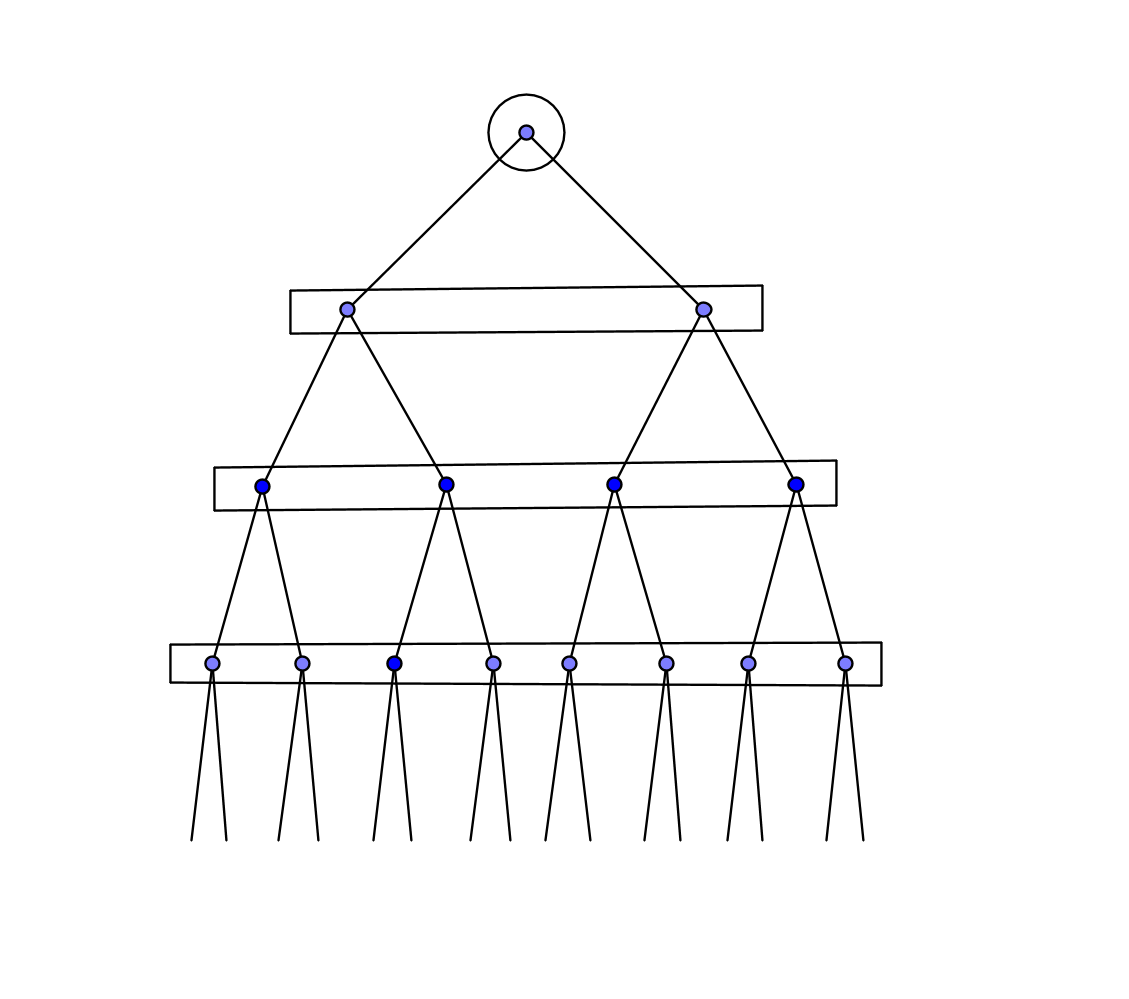}

\vspace*{-0.7cm}

\caption{Spheres $S_r$ in the case of a binary tree graph, where $k = 2$.}
\end{center}
\end{figure}

\noindent
In this case, we have
\begin{equation}
\displaystyle \mathcal{V} = \bigsqcup_{0}^\infty S_r,
\end{equation}
where $\bigsqcup$ means a disjoint union, so that we have
\begin{equation}
\ell^2(\mathcal{V}) = \bigoplus_0^\infty \ell^2(S_r).
\end{equation}
In this paper, we are interested in the case of regular rooted tree graphs \eqref{eq1,60} 
with $k$-fold branching, $k \geq 1$. Moreover, the potential $M$ will be assumed to satisfy 
the following assumption:

\medskip

\noindent
{\bf Assumption (A)}: For $ v \in \mathcal{V}$, we have
\begin{equation}\label{eq1,7}
\vert M(v)\vert \leq {\rm Const.} \, e^{-\delta \vert v \vert}, \quad {\rm with} \quad
\begin{cases} 
\delta > 0 & \text{if } \, k = 1, \\ 
\delta \geq 6 \ln (k) & {\rm otherwise}.
\end{cases}
\end{equation}

\begin{rem}\label{r1,1}
We point out that in Assumption (A) above, there is no restriction on the 
perturbation potential $M$ concerning its self-adjointness or not.
The case $k = 1$ includes in particular the case of the Laplacian on 
$\ell^2(\bn,\bc)$ without any boundary condition at $0$.
\end{rem}

As mentioned above, in this article we investigate the resonances (or eigenvalues) 
distribution for the operator $\ch_M$ near the spectral thresholds $t_\pm(k)$ given by 
\eqref{eq1,61}. As this will be observed, the work of Allard and Froese \cite{af} will 
play an important role in our analysis (cf. Section \ref{sec3} for more details). 
More precisely, in order to establish a suitable representation of the resolvent 
associated to the operator $-L + k + 1$, $k \geq 1$ (cf. Theorem \ref{p4,1}). 
Under Assumption (A), the perturbation potential $\widetilde{M}$ satisfies the 
decay assumption of $M$. So, if we let $\Lambda_n$ to denote the orthogonal projection 
onto $\bigoplus_{r=0}^n \ell^2(S_r)$, then with the aid of the Schur lemma, it can
be shown that 
\begin{equation}\label{eq,l}
\Vert \widetilde{M} - \widetilde{M} \Lambda_n \Vert \underset{n \rightarrow 
\infty}{\longrightarrow} 0.
\end{equation}
Since $\ell^2(S_r)$ is a $k^r$-dimensional space, then \eqref{eq,l} implies that 
the operator 
$\widetilde{M}$ is the limit in norm of a sequence of finite rank operators. 
Therefore, $\widetilde{M}$ is a compact operator and in particular it is relatively 
compact with respect to the operator $-L + k + 1$. Thus, since the operator 
$-L + k + 1$ is self-adjoint, then by \cite[Theorem 2.1, p. 373]{gohs} we have a 
disjoint union
\begin{equation}
\sigma(\ch_{\widetilde M}) = \sigma_{\text{\bf ess}}(\ch_{\widetilde M}) \bigsqcup 
\sigma_{\text{\bf disc}}(\ch_{\widetilde M}).
\end{equation} 
Moreover, Weyl's criterion on the invariance of the essential spectrum implies that
\begin{equation}
\sigma_{\textup{\bf ess}} (\ch_{\widetilde M}) = \sigma_{\textup{\bf ess}} 
(-L + k + 1) = \left[ t_-(k),t_+(k) \right].
\end{equation}
However, the (complex) discrete spectrum $\sigma_{\text{\bf disc}}(\ch_{\widetilde M})$
generated by the potential $\widetilde{M}$ can only accumulate at the points of 
$\sigma_{\textup{\bf ess}} (\ch_{\widetilde M})$.

\begin{rem}
When ${\widetilde M} = {\widetilde M}^\ast$, 
$\sigma_{\textup{\bf disc}} (\ch_{\widetilde M})$ is just the set of real eigenvalues 
of $\ch_{\widetilde M}$ respectively from the right and the left of $t_\pm(k)$. 
\end{rem}

\noindent
Exploiting the exponential decay of the potential $\widetilde{M}$, we extend 
(cf. Section \ref{sec5}) meromorphically in Banach weighted spaces the resolvent of the 
operator $\ch_{\widetilde M}$ near $z = t_\pm(k)$, in some two sheets Riemann surfaces 
$\mathcal{M}_{t_\pm}$ respectively. The first main difficulty to overcome 
is to establish a good representation of the kernel of the resolvent associated to the
operator $-L + k + 1$, $k \ge 1$ (cf. Section \ref{sec4}). We thus define the resonances of 
the operator $\ch_{\widetilde M}$ near $z = t_\pm(k)$ as the poles of the above 
meromorphic extensions. Notice that this set of resonances contains the eigenvalues 
of the operator $\ch_{\widetilde M}$ localized near the spectral thresholds $t_\pm(k)$. 
Otherwise, in the two sheets Riemann surfaces $\mathcal{M}_{t_\pm}$,
the resonances will be parametrized by $z_{t_\pm}(\lambda)$ for $\lambda$ 
sufficiently small for technical reasons. Furthermore, the point $\lambda = 0$ corresponds to the threshold 
$z = t_\pm(k)$ (cf. Section \ref{sec5} for more details). Actually, the resonances 
verifying
\begin{equation}
z_{t_\pm}(\lambda) \in \mathcal{M}_{t_\pm}, \quad \im(\lambda) < 0,
\end{equation}
live in the non physical plane while those verifying 
\begin{equation}
z_{t_\pm}(\lambda) \in \mathcal{M}_{t_\pm}, \quad \im(\lambda) \geq 0,
\end{equation} 
coincide with the discrete and the embedded eigenvalues of the operator $\ch_{\widetilde M}$ 
near $t_\pm$ and are localized in the physical plane. 
We state Theorem \ref{t1} where we establish an absence of resonances of the operator 
$\ch_{\widetilde M}$ near the spectral thresholds $t_\pm(k)$. In particular, this implies 
results on the absence of discrete spectrum and embedded eigenvalues near $t_\pm(k)$ 
(cf. Corollaries \ref{t2} and \ref{t3}). To prove these results, we first reduce the analysis 
of resonances near the thresholds $t_\pm(k)$ to that of the noninvertibility of some
nonself-adjoint compact operators near $\lambda = 0$ (cf. Propositions \ref{p5,2} and 
\ref{p5,3}). This can be seen as a Birman-Schwinger principle in a nonself-adjoint context.
Afterwards, the reduction made on the problem is reformulated in terms of characteristic
values problems (cf. Propositions \ref{p6,2} and \ref{p6,3}). This allows us to apply 
powerful results (cf. Section \ref{sa}) on the theory of characteristic 
values of finite meromorphic operator-valued functions to conclude.

\section{Statement of the main results}\label{secp}

Let us first fix some notations. If $\lambda \in \bc$, as usual 
$\vert \lambda \vert <\!\!<1$ means that $\lambda$ is chosen small enough. The set 
of resonances of the operator $\ch_{\widetilde M}$ near the spectral thresholds 
$t_\pm(k)$ given by \eqref{eq1,61} will be respectively denoted by
\begin{equation}
{\rm Res}_{t_\pm} \, (\ch_{\widetilde M}).
\end{equation}
We also recall that near $z = t_\pm(k)$, the 
resonances are defined in some Riemann surfaces $\mathcal{M}_{t_\pm}$ and coincide 
with $z_{t_\pm}(\lambda)$, $0 < \vert \lambda \vert <\!\!<1 $. More precisely, 
they are parametrized respectively by
\begin{equation}\label{p1}
z_{t_\pm}(\lambda) := t_\pm(k) \mp \lambda^2\sqrt{k} \in \mathcal{M}_{t_\pm}.
\end{equation}
Furthermore, the embedded eigenvalues and the discrete spectrum of the operator 
$\ch_{\widetilde M}$ near $t_\pm(k)$ are the resonances 
$z_{t_\pm}(\lambda) \in \mathcal{M}_{t_\pm}$ with $\im(\lambda) \geq 0$. 

\medskip

\noindent
Now, for $0 < r <\!\!<1$, let us introduce the punctured neighborhood of $\lambda = 0$
\begin{equation}
\Omega_r^\ast := \big \lbrace \lambda \in \bc : 0 < \vert \lambda \vert < r \big \rbrace.
\end{equation}
We then can state our first main result that gives an absence of resonances of 
the operator $\ch_{\widetilde M}$ near the thresholds $t_\pm(k)$, in small domains 
of the form $t_\pm(k) \mp \sqrt{k} \, {\Omega_r^\ast}^2$.

\begin{theo}[\textbf{Absence of resonances}]\label{t1}
Assume that the potential $M$ satisfies Assumption (A). Then, for any $r > 0$ small
enough and any punctured neighborhood $\Omega_r^\ast$, we have
\begin{equation}
\# \big \lbrace z_{t_\pm}(\lambda) \in Res_{t_\pm}(\ch_{\widetilde M}) : \lambda \in 
\Omega_r^\ast \big \rbrace = 0,
\end{equation}
the resonances being counted with their multiplicity given by 
\eqref{eq5,25} and \eqref{eq5,31}.
\end{theo}

\noindent
Notice that Theorem \ref{t1} just says that the operator $\ch_{\widetilde M}$ has no 
resonances in a punctured neighborhood of $t_\pm(k)$ in the two-sheets Riemann surfaces $\mathcal{M}_{t_\pm}$ where they are defined.

\begin{figure}[h]\label{fig 1}
\begin{center}
\tikzstyle{+grisEncadre}=[fill=gray!60]
\tikzstyle{blancEncadre}=[fill=white!100]
\tikzstyle{grisEncadre}=[fill=gray!20]
\begin{tikzpicture}[scale=1.2]

\draw [grisEncadre] (0,0) -- (180:1.58) arc (180:360:1.58) -- cycle;
\draw [blancEncadre] (0,0) -- (0:1.58) arc (0:180:1.58) -- cycle;


\draw (0,0) circle (0.5);
\draw [->] [thick] (-2.5,0) -- (2.3,0);
\draw (2.3,0) node[right] {\small{$\re(\lambda)$}};
\draw [->] [thick] (0,-2.5) -- (0,2.7);
\draw (0.05,2.7) node[right] {\small{$\im(\lambda)$}};
\draw (0.26,-0.40) node[above] {\tiny{$r$}};
\draw (0,0) -- (0.24,-0.45);
\draw (0,0) -- (1.31,0.9);
\draw (0.8,0.55) node[above] {\tiny{$r_{0}$}};

\node at (-1.4,-1.2) {\tiny{$\times$}};

\node at (0.9,-1.4) {\tiny{$\times$}};
\node at (1.8,-0.2) {\tiny{$\times$}};
\node at (1.4,-1.2) {\tiny{$\times$}};

\node at (0.9,1.4) {\tiny{$\times$}};
\node at (1.4,1.2) {\tiny{$\times$}};

\node at (-1.7,0.4) {\tiny{$\times$}};
\node at (-1.4,1.2) {\tiny{$\times$}};

\node at (-0.5,1.7) {\tiny{$\times$}};

\node at (-0.2,1.7) {\tiny{$\times$}};

\node at (2,2.3) {\small{$\textup{\textbf{Absence of resonances}}$}};
\draw [->] [dotted] (1.8,2.15) -- (-0.5,1);
\draw [->] [dotted] (1.8,2.15) -- (1,-0.5);

\node at (-4.5,0.5) {\small{$\textup{Corresponds to the \textbf{Pysical plane}}$}};
\node at (-4.5,0.2) {\small{$\textup{(in variable $z$)}$}};
\draw [->] [dotted] (-2.1,0.5) -- (-1,0.5);

\node at (-4.67,-0.5) {\small{$\textup{Corresponds to the \textbf{Non pysical}}$}};
\node at (-4.55,-0.8) {\small{$\textup{\textbf{plane}}$}};
\draw [->] [dotted] (-1.8,-0.5) -- (-1,-0.5);

\end{tikzpicture}
\caption{\textup{\textbf{Resonances near $t_\pm(k)$ in variable 
$\lambda$:} For $r_0$ sufficiently small, thanks to Theorem \ref{t1}, the operator 
$\ch_{\widetilde M}$ has no resonances $z_{t_\pm}(\lambda)$ in a vicinity of 
$t_\pm(k)$ in $\big\lbrace \lambda : r < \vert \lambda \vert < r_0 \big\rbrace$, 
$r \searrow 0$.}}
\end{center}
\end{figure}

\medskip

\noindent
Since near $t_\pm(k)$ the discrete spectrum of the operator $\ch_{\widetilde M}$ 
corresponds to resonance points $z_{t_\pm}(\lambda) \in \mathcal{M}_{t_\pm}$ 
with $\im(\lambda) \geq 0$, then a first consequence of Theorem \ref{t1} is the 
following result giving a non cluster phenomena of real or non real eigenvalues 
near $t_\pm(k)$.

\begin{cor}[\textbf{Non cluster of eigenvalues}]\label{t2}
Assume that the potential $M$ satisfies Assumption (A). Then, there is no sequence 
$(\nu_j)_j$ of non real or real eigenvalues of the operator $\ch_{\widetilde M}$ 
accumulating at $t_\pm(k)$.
\end{cor}

\noindent
Now, thanks to the parametrizations \eqref{p1}, the embedded eigenvalues of the 
operator $\ch_{\widetilde M}$ near $t_\pm(k)$ respectively from the left and the 
right are the resonances $z_{t_\pm}(\lambda) = t_\pm(k) \mp \lambda^2\sqrt{k}$ 
with $\lambda \in \br_+$ sufficiently small. Therefore, as a second consequence of 
Theorem \ref{t1} together with \cite[Theorem 9]{af}, we have the following:

\begin{cor}[\textbf{Absence of embedded eigenvalues}]\label{t3}
Assume that the potential $M$ satisfies Assumption (A). Then, for any $r > 0$ small
enough, the operator $\ch_{\widetilde M}$ has no embedded eigenvalues in 
\begin{equation}
\left( t_-(k),t_-(k) + r^2 \right) \cup \left( t_+(k) - r^2,t_+(k) \right).
\end{equation} 
In particular, for $M = M^\ast$, the set of embedded eigenvalues of the operator 
$\ch_{\widetilde M}$ in 
$\left( t_-(k),t_+(k) \right)$ is finite. 
\end{cor}

\begin{rem}
\textup{Note that our results can be extended to the case of Bethe or Cayley
graphs.}
\end{rem}

\section{On a diagonalization of the operator $L$ for a regular tree graph}\label{sec3}

In this section, we summarize some results and tools we need and which are developed in \cite{af,al}. We shall essentially follow \cite[Section 3]{af}
and we refer to the cited papers for more details.

\smallskip

Define the operator $\Pi$ on $\ell^2(\mathcal{V})$ by
\begin{equation}\label{eq3,1}
(\Pi \phi)(v) := \sum_{w \, : \, w \rightarrow v} \phi(w),
\end{equation}
where for two vertices $v$ and $w$, $v \rightarrow w$ means that they are connected
by an edge with $\vert w \vert = \vert v \vert + 1$. Using the inner product defined
on the Hilbert space $\ell^2(\mathcal{V})$ by \eqref{eq,ps}, it can be easily checked 
that the adjoint operator $\Pi^\ast$ is given by
\begin{equation}\label{eq3,2}
(\Pi^\ast \phi)(v) := \sum_{w \, : \, v \rightarrow w} \phi(w).
\end{equation}
If we let $L_S$ be the spherical Laplacian defined on $\ell^2(\mathcal{V})$ by
\begin{equation}\label{eq3,3}
(L_S \phi)(v) := \sum_{\substack{w \, : \, w \sim v \\ \vert w \vert = \vert v \vert}} \phi(w),
\end{equation}
then the operator $L$ given by \eqref{eq1,2}
can be written as
\begin{equation}\label{eq3,4}
L = \Pi + \Pi^\ast + L_S.
\end{equation}
In a regular rooted tree graph with $k$-fold branching, since there are 
no edges connecting vertices within each sphere, then $L_S = 0$ so that
\begin{equation}\label{eq3,5}
L = \Pi + \Pi^\ast.
\end{equation}
To diagonalize the operator $L$ given by \eqref{eq3,5}, invariant subspaces $M_n$, 
$n \geq 0$, for $\Pi$ are firstly constructed in \cite{af}. More precisely, we have the 
following lemma:

\begin{lem}{\cite[Lemma 1]{af}}\label{l3,1}
The Hilbert space $\ell^2(\mathcal{V})$ can be decomposed as an orthogonal direct sum
\begin{equation}\label{eq3,6}
\ell^2(\mathcal{V}) = \bigoplus_{n = 0}^\infty M_n = 
\bigoplus_{n = 0}^\infty \bigoplus_{r = n}^\infty Q_{n,r},
\end{equation}
where the subspaces $M_n$ are  $L$-invariant.
\end{lem}

\noindent
By construction in this Lemma \ref{l3,1}, we have $Q_{0,0} := \ell^2(S_0)$ and  
$\ell^2(S_r) = \bigoplus_{\ell = 0}^r Q_{\ell,r}$. A schematic interpretation
yields a triangular diagram as in Figure 4.1 below.


\begin{figure}[h]\label{fig 2}
\begin{center}

\includegraphics[scale=1]{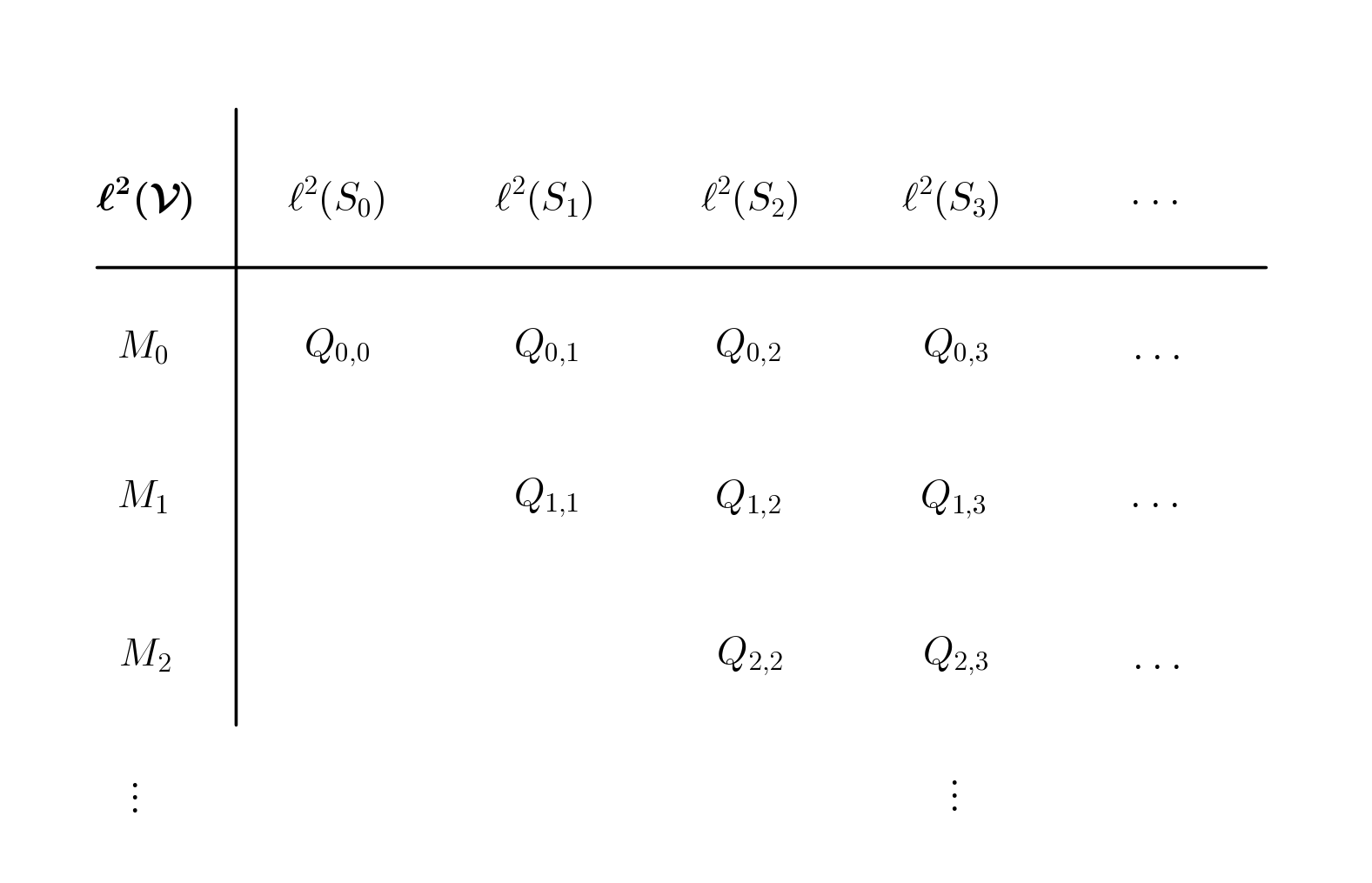}

\vspace*{-0.5cm}

\caption{Orthogonal subspace decomposition}
\end{center}
\end{figure}

\medskip

\noindent
According to Lemma \ref{l3,1}, for any $n \geq 0$, the subspace $M_n$ is invariant for the 
operator $L$. Thus, $L$ can be decomposed as
\begin{equation}\label{eq3,60}
L = \bigoplus_{n = 0}^{\infty} L_n,
\end{equation}
the operators $L_n$, $n \geq 0$, being the restriction of $L$ to $M_n$. So, in order to 
diagonalize the operator $L$, it suffices to do it for each operator $L_n$ for $n \geq 0$. 
Consider a vector $\phi \in M_n$, i.e.
\begin{equation}\label{eq3,7}
\phi = \displaystyle\oplus_{j = 0}^{\infty} \phi_{n,n+j},
\end{equation}
with $\phi_{n,n+j} \in Q_{n,n+j}$ for any $j \geq 0$. The idea is to construct an
isomorphism between the subspace $M_n$ and the space of $Q_{n,n}$-valued sequences, 
namely the space $\ell^2(\bz^+,Q_{n,n})$. By construction of the $Q_{n,r}$, $n \geq 0$,
$r \geq 0$ (see for instance \cite{af}), for any $j \geq 0$, the operator 
$\big( \frac{1}{\sqrt{k}} \Pi \big)^j$ defines an isometry between $Q_{n,n+j}$ and
$Q_{n,n}$, and \eqref{eq3,7} can be written as
\begin{equation}\label{eq3,8}
\phi = \oplus_{j = 0}^{\infty} \left( \frac{1}{\sqrt{k}} \Pi \right)^j \chi_{n + j},
\end{equation}
where $\chi := (\chi_{n + j})_{j \geq 0}$ defines a sequence of vectors lying in 
$Q_{n,n}$. Therefore, under the above considerations, the operator
\begin{equation}\label{eq3,9}
W : M_n \longrightarrow \ell^2(\bz^+,Q_{n,n}), \quad W \phi = \chi,
\end{equation}
defines an isomorphism between the spaces $M_n$ and $\ell^2(\bz^+,Q_{n,n})$. Indeed, 
we have
\begin{equation}\label{eq3,90}
\langle \phi,\phi \rangle = \sum_{j = 0}^{\infty} \langle \phi_{n + j},\phi_{n + j} 
\rangle = \sum_{j = 0}^{\infty} \langle \chi_{n + j},\chi_{n + j} 
\rangle = \langle \chi,\chi \rangle_{\ell^2(\bz^+,Q_{n,n})}.
\end{equation}
Now, let $\bt := \br /2\pi \bz$ be the torus and define the unitary operator 
\begin{equation}\label{eq3,10}
U : M_n \cong \ell^2(\bz^+,Q_{n,n}) \longrightarrow L^2_{\rm odd} (\mathbb{T},Q_{n,n})
\end{equation}
acting as
\begin{equation}\label{eq3,11}
U \big( (\chi_n,\chi_{n + 1}, \ldots ) \big) := \frac{1}{\sqrt{\pi}} \sum_{j = 0}^\infty
\chi_{n + j} \sin \big( (j + 1) \theta \big).
\end{equation}
Notice that the inner product in $L^2_{\rm odd} (\mathbb{T},Q_{n,n})$ is defined by
\begin{equation}\label{eq3,12}
\langle f,g \rangle_{L^2_{\rm odd}} := \int_{\mathbb{T}} \langle f(\theta),g(\theta) 
\rangle \, d\theta.
\end{equation}
Hence, a direct computation shows that 
\begin{equation}\label{eq3,13}
\langle U \chi,U \chi \rangle_{L^2_{\rm odd}} = 
\langle \chi,\chi \rangle_{\ell^2(\bz^+,Q_{n,n})} = \langle \phi,\phi \rangle,
\end{equation}
where the last equality corresponds to \eqref{eq3,90}. Moreover, we have the following lemma:

\begin{lem}{\cite[Lemma 2]{af}}\label{l3,2}
For any $n \geq 0$, we have
\begin{equation}\label{eq3,14}
U L_n U^\ast = 2\sqrt{k} \cos (\theta).
\end{equation}
\end{lem}

\noindent
In particular, Lemma \ref{l3,2} shows that for any $n \geq 0$, the spectrum of the 
operator $L_n$ is equal to $\big[ -2\sqrt{k},2\sqrt{k} \big]$ and is absolutely 
continuous. Hence, this implies \eqref{eq1,4}.

\section{Representation of the weighted resolvent $A (-L + k + 1 - z)^{-1} B^\ast$}\label{sec4}

In this section, we give a suitable representation of the weighted resolvent 
$A (-L + k + 1 - z)^{-1} B^\ast$ which turns to be useful in our analysis, where $A$ 
and $B$ are bounded operators on $\ell^2(\mathcal{V})$.

\medskip

For $n \geq 0$, let $P_n$ be the orthogonal projection of $\ell^2(\mathcal{V})$ 
onto $M_n$, the subspace defined in Lemma \ref{l3,1}. Since $M_n$ can be decomposed as
\begin{equation}\label{eq4,1}
M_n = \bigoplus_{j = 0}^{\infty} Q_{n,n + j},
\end{equation}
then if we let $(E_m^{n,n+j})_{0 \leq m \leq N_j}$ denote an orthonormal basis of the
finite-dimensional space $Q_{n,n + j}$ for any $j$ fixed, we have
\begin{equation}\label{eq4,2}
P_n = \sum_{j \, \geq \, 0} \sum_{0 \leq m \leq N_j} \langle \cdot,E_m^{n,n+j} \rangle 
E_m^{n,n+j}.
\end{equation}
Notice that for any $j \geq 0$ fixed, we have 
\begin{equation}\label{eq4,3}
1 + N_j = \dim \, Q_{n,n+j} < k^{n + j} = \dim \, \ell^2(S_{n + j}),
\end{equation}
since $Q_{n,n+j} \subset \ell^2(S_{n + j})$. Furthermore, according to Section 
\ref{sec3}, for any $0 \leq m \leq N_j$ fixed, there exists a unique vector 
$\chi_m^{n,n+j} \in Q_{n,n}$ such that
\begin{equation}\label{eq4,4}
E_m^{n,n+j} = \left( \frac{1}{\sqrt{k}} \Pi \right)^j \chi_m^{n,n + j}.
\end{equation}
Our goal in this section is to prove the following result:

\begin{theo}\label{p4,1}
Let $A$ and $B$ be two bounded operators on $\ell^2(\mathcal{V})$. Then, for any $z$ 
in the resolvent set of the operator $-L + k + 1$ and any $\varphi \in \ell^2(\mathcal{V})$, 
we have
\begin{equation}\label{eq4,5}
A (-L + k + 1 - z)^{-1} B^\ast 
\varphi (v) = \sum_{v' \in \mathcal{V}} K(v,v') \varphi (v'),
\end{equation}
where the kernel $K(v,v')$ is given by
\begin{equation}\label{eq4,6}
\begin{split}
K(v,v') := & \frac{1}{2 \sqrt{k}} \sqrt{\frac{2}{\pi}} \sum_{n \, \geq \, 0} 
\sum_{\substack{j \, \geq \, 0 \\ \ell \, \geq \, 0}} 
\sum_{\substack{0 \leq m \leq N_j \\ 0 \leq q \leq N_\ell}} 
\overline{ \Big( B E_m^{n,n+j} \Big) (v')}
\langle \chi_m^{n,n + j},\chi_q^{n,n + \ell} \rangle \\
& \times \left( - \frac{i e^{i(j + \ell + 2)\Phi}}{\sqrt{u}
\sqrt{4 - u}} + \frac{i e^{i\vert j - \ell \vert\Phi}}{\sqrt{u} \sqrt{4 - u}} \right)
\Big( A E_q^{n,n + \ell} \Big) (v) ,
\end{split}
\end{equation}
throughout the double change of variables
\begin{equation}\label{eq4,60}
\frac{z + 2\sqrt{k} - (k + 1)}{\sqrt{k}} = u = 4 \sin^2 \left( \frac{\Phi}{2} \right), 
\quad \im (\Phi) > 0.
\end{equation}
\end{theo}

\noindent
\begin{prof}
Let $\varphi \in \ell^2(\mathcal{V})$ and $z \in \rho (-L + k + 1)$ the resolvent set of 
the operator $-L + k + 1$. Thanks to \eqref{eq3,60}, we have
\begin{equation}\label{eq4,7}
\begin{split}
(-L + k + 1 - z)^{-1} & = \sum_{n \, \geq \, 0} P_n (-L_n + k + 1 - z)^{-1} P_n \\
& = \sum_{n \, \geq \, 0} P_n U^\ast U (-L_n + k + 1 - z)^{-1} U^\ast U P_n,
\end{split}
\end{equation}
where $U$ is the unitary operator defined by \eqref{eq3,10}. Thus, for any vector
$\psi \in \ell^2(\mathcal{V})$, we have
\begin{equation}\label{eq4,8}
\begin{split}
& \biggl\langle  A (-L + k + 1 - z)^{-1} B^\ast 
\varphi,\psi \biggr\rangle \\
& = \sum_{n \, \geq \, 0} \biggl\langle U (-L_n + k + 1 - z)^{-1} U^\ast U P_n
B^\ast \varphi,U P_n A^\ast \psi 
\biggr\rangle_{L^2_{\rm odd}},
\end{split}
\end{equation}
$\langle \cdot,\cdot \rangle_{L^2_{\rm odd}}$ being the inner product defined by
\eqref{eq3,12}. Together with Lemma \ref{l3,2}, this gives
\begin{equation}\label{eq4,9}
\begin{split}
& \biggl\langle A (-L + k + 1 - z)^{-1} B^\ast
\varphi,\psi \biggr\rangle \\
& = \sum_{n \, \geq \, 0} \int_{\mathbb{T}} \big( -2\sqrt{k} \cos (\theta) + k + 1 - z \big)^{-1} 
\biggl\langle U P_n B^\ast \varphi (\theta),U P_n 
A^\ast \psi (\theta) \biggr\rangle \, d\theta.
\end{split}
\end{equation}
From \eqref{eq4,2}, it follows that for any $\phi \in \ell^2(\mathcal{V})$ and any bounded
operator $W$ on $\ell^2(\mathcal{V})$, we have
\begin{equation}\label{eq4,10}
U P_n W \phi (\theta) = \sum_{j \, \geq \, 0} 
\sum_{0 \leq m \leq N_j} \biggl\langle \phi,W^\ast E_m^{n,n+j} 
\biggr\rangle U E_m^{n,n+j} (\theta).
\end{equation}
Then, combining \eqref{eq4,9} and \eqref{eq4,10}, we obtain
\begin{equation}\label{eq4,11}
\begin{split}
 \biggl\langle & A (-L + k + 1 - z)^{-1} B^\ast
\varphi,\psi \biggr\rangle \\
& = \sum_{n \, \geq \, 0} \sum_{\substack{j \, \geq \, 0 \\ \ell \, \geq \, 0}} 
\sum_{\substack{0 \leq m \leq N_j \\ 0 \leq q \leq N_\ell}} 
\biggl\langle \varphi,B E_m^{n,n+j} 
\biggr\rangle \overline{\biggl\langle \psi,A E_q^{n,n+\ell} 
\biggr\rangle} \\
& \times \int_{\mathbb{T}} \big( -2\sqrt{k} \cos (\theta) + k + 1 - z \big)^{-1} 
\biggl\langle U E_m^{n,n+j} (\theta),U E_q^{n,n+\ell} (\theta) 
\biggr\rangle \, d\theta \\
& = \sum_{n \, \geq \, 0} \sum_{\substack{j \, \geq \, 0 \\ \ell \, \geq \, 0}} 
\sum_{\substack{0 \leq m \leq N_j \\ 0 \leq q \leq N_\ell}} 
\biggl\langle \biggl\langle \varphi,B E_m^{n,n+j} \biggr\rangle 
A E_q^{n,n+\ell},\psi \biggr\rangle \\
& \times \int_{\mathbb{T}} \big( -2\sqrt{k} \cos (\theta) + k + 1 - z \big)^{-1} 
\biggl\langle U E_m^{n,n+j} (\theta),U E_q^{n,n+\ell} (\theta) 
\biggr\rangle \, d\theta.
\end{split}
\end{equation}
According to the construction of the unitary operator $U$ and \eqref{eq4,4}, for 
$p = m$, $q$ and $p' = j$, $\ell$ respectively, we have
\begin{equation}
U E_p^{n,n+p'} (\theta) = \frac{1}{\sqrt{\pi}} \chi_p^{n,n+p'} 
\sin \big( (p' + 1) \theta \big).
\end{equation}
Putting this together with \eqref{eq4,11}, we obtain
\begin{equation}\label{eq4,12}
\begin{split}
& \biggl\langle A (-L + k + 1 - z)^{-1} B^\ast 
\varphi,\psi \biggr\rangle \\
& = \frac{1}{\pi} \sum_{n \, \geq \, 0} \sum_{\substack{j \, \geq \, 0 \\ \ell \, \geq \, 0}} 
\sum_{\substack{0 \leq m \leq N_j \\ 0 \leq q \leq N_\ell}} 
\biggl\langle \biggl\langle \varphi,B E_m^{n,n+j} \biggr\rangle 
\langle \chi_m^{n,n+j},\chi_q^{n,n+\ell} \rangle A E_q^{n,n+\ell},\psi \biggr\rangle \\
& \times \int_{\mathbb{T}} \big( -2\sqrt{k} \cos (\theta) + k + 1 - z \big)^{-1} 
\sin \big( (j + 1) \theta \big) \sin \big( (\ell + 1) \theta \big) \, d\theta.
\end{split}
\end{equation}
Now, \eqref{eq4,12} implies that the action of the operator  
$A (-L + k + 1 - z)^{-1} B^\ast$ on $\ell^2(\mathcal{V})$
can be described by
\begin{equation}\label{eq4,13}
\begin{split}
& A (-L + k + 1 - z)^{-1} B^\ast \varphi (v) \\
& = \frac{1}{\pi} \sum_{n \, \geq \, 0} \sum_{\substack{j \, \geq \, 0 \\ \ell \, \geq \, 0}} 
\sum_{\substack{0 \leq m \leq N_j \\ 0 \leq q \leq N_\ell}} 
\biggl\langle \varphi,B E_m^{n,n+j} \biggr\rangle 
\Big( A E_q^{n,n+\ell} \Big) (v) 
\langle \chi_m^{n,n+j},\chi_q^{n,n+\ell} \rangle \\
& \times \int_{\mathbb{T}} \big( -2\sqrt{k} \cos (\theta) + k + 1 - z \big)^{-1} 
\sin \big( (j + 1) \theta \big) \sin \big( (\ell + 1) \theta \big) \, d\theta.
\end{split}
\end{equation}
Since we have
\begin{equation}
\biggl\langle \varphi,W E_m^{n,n+j} \biggr\rangle = 
\sum_{v' \in \mathcal{V}} \varphi (v') \overline{ \Big( W E_m^{n,n+j} \Big) (v')},
\end{equation}
$W$ being as above, then it follows from \eqref{eq4,13} that
\begin{equation}
A (-L + k + 1 - z)^{-1} B^\ast 
\varphi (v) = \sum_{v' \in \mathcal{V}} K(v,v') \varphi (v'),
\end{equation}
with
\begin{equation}\label{eq4,14}
\begin{split}
 K(v,v') :=  
& \frac{1}{\pi} \sum_{n \, \geq \, 0} \sum_{\substack{j \, \geq \, 0 \\ \ell \, \geq \, 0}} 
\sum_{\substack{0 \leq m \leq N_j \\ 0 \leq q \leq N_\ell}} 
\overline{ \Big( B E_m^{n,n+j} \Big) (v')} 
\Big( A E_q^{n,n+\ell} \Big) (v) 
\langle \chi_m^{n,n+j},\chi_q^{n,n+\ell} \rangle \\
& \times \int_{\mathbb{T}} \big( -2\sqrt{k} \cos (\theta) + k + 1 - z \big)^{-1} 
\sin \big( (j + 1) \theta \big) \sin \big( (\ell + 1) \theta \big) \, d\theta.
\end{split}
\end{equation}
Then, to complete the proof of the theorem, it remains only to show that
\begin{equation}\label{eq4,15}
\begin{split}
\frac{1}{\pi} \int_{\mathbb{T}} & \big( -2\sqrt{k} \cos (\theta) + k + 1 - z \big)^{-1} 
\sin \big( (j + 1) \theta \big) \sin \big( (\ell + 1) \theta \big) \, d\theta \\
& = \frac{1}{2\sqrt{k}} \sqrt{\frac{2}{\pi}} \left( - \frac{i e^{i(j + \ell + 2)\Phi}}{\sqrt{u}
\sqrt{4 - u}} + \frac{i e^{i\vert j - \ell \vert\Phi}}{\sqrt{u} \sqrt{4 - u}} \right),
\end{split}
\end{equation}
where the relation between $z$, $u$ and $\Phi$ is given by \eqref{eq4,60}. To do this, 
we have to deal with the discrete Fourier transform $\mathcal{F} : \ell^2(\bz,\bc) 
\rightarrow L^2(\bt)$, defined for any $x \in \ell^2(\bz,\bc)$ and $f \in L^2(\bt)$ by
\begin{equation}
(\mathcal{F}x)(\theta) := (2\pi)^{-\frac{1}{2}} \sum_{n \, \in \, \bz}
e^{-in\theta} x(n), \quad
\big( \mathcal{F}^{-1} f \big)(n) := (2\pi)^{-\frac{1}{2}} 
\int_{\bt} e^{in\theta} f(\theta) \, d\theta.
\end{equation}
Let $u \in \bc \setminus [0,4]$ and introduce the following change of variables
\begin{equation}
u = 4 \sin^2 \left( \frac{\Phi}{2} \right), \quad \im(\Phi) > 0.
\end{equation}
Then, it can be proved (cf. e.g. \cite[Section 2]{it}) that we have
\begin{equation}
\Big( \mathcal{F}^{-1} \big( 2 - 2 \cos (\cdot) - u \big)^{-1} \Big)(n)
= \frac{ie^{i \vert n \vert \Phi}}{2 \sin(\Phi)} = 
\frac{ie^{i \vert n \vert \Phi}}{\sqrt{u} \sqrt{4 - u}},
\end{equation}
or equivalently
\begin{equation}\label{eq4,16}
\frac{1}{\pi} \int_{\bt} \big( 2 - 2 \cos(\theta) - u \big)^{-1} e^{in\theta} \, d\theta
= \sqrt{\frac{2}{\pi}} \frac{ie^{i \vert n \vert \Phi}}{\sqrt{u} \sqrt{4 - u}}.
\end{equation}
Now, \eqref{eq4,16} with the help of the transformations 
$\sin(n\theta) = \frac{1}{2i} \left( e^{in\theta} - e^{-in\theta} \right)$ and
\begin{equation}
\Big( -2\sqrt{k} \cos(\theta) + k + 1 - z \Big)^{-1} = \frac{1}{\sqrt{k}}
\Big( 2 - 2 \cos(\theta) - u \Big)^{-1},
\end{equation}
where $u = \frac{z + 2\sqrt{k} - (k + 1)}{\sqrt{k}}$ give immediately \eqref{eq4,15}.
This completes the proof of the theorem.
\end{prof}

\section{Resonances near $z = t_\pm(k)$}\label{sec5}

\subsection{Definition of the resonances}

In this subsection, we define the resonances of the operator $\ch_{\widetilde M}$ 
near the spectral thresholds $t_\pm(k)$ given by \eqref{eq1,60}. As preparation, 
preliminary lemmas will be proved firstly.

\medskip

From now on, the potential perturbation $M$ is assumed to satisfy Assumption (A). 
Moreover, the following determination of the complex square root
\begin{equation}\label{eq5,1}
\bc \setminus (-\infty,0] \overset{\sqrt{\cdot}}
{\longrightarrow} \bc^+ := \big\lbrace z \in \bc : \im(z) > 0 \big \rbrace
\end{equation}
will be adopted throughout this paper. For $\varepsilon > 0$ such that 
$0 < \varepsilon < \frac{\delta}{4}$, we let $D(0,\varepsilon)^\ast$
be the punctured neighborhood of $0$ defined by
\begin{equation}\label{eq5,2}
D(0,\varepsilon)^\ast := \big \lbrace \lambda \in \bc : 0 < \vert \lambda \vert < 
\varepsilon \big \rbrace.
\end{equation}
Thanks to the first change of variables in \eqref{eq4,60}, to define and to study the 
resonances of the operator $\ch_{\widetilde M}$ near the spectral thresholds $t_\pm(k)$, 
it suffices to define and to study them respectively near $u = 0$ and $u = 4$. However, 
in practice, there is a simple way (see the comments just after Definition \ref{d5,1}) 
allowing to reduce the analysis of the resonances near the second threshold $t_+(k)$ to 
that of the first one $t_-(k)$. For further use, let $e_{\pm}$ be the multiplication 
operators by the functions
\begin{equation}\label{eq5,20}
v \longmapsto e_\pm (v) := e^{\pm\frac{\delta}{2}\vert v \vert}.
\end{equation}
We have the following lemma:

\begin{lem}\label{l5,1}
Let $z_{t_-}(\lambda)$ be the parametrization defined by \eqref{p1}. Then, there exists 
$0 < \varepsilon_0 \le \frac{\delta}{8}$ small enough such that the operator-valued function
\begin{equation}\label{eq5,3}
\lambda \mapsto e_- \big( -L + k + 1 - z_{t_-}(\lambda) \big)^{-1} e_-,
\end{equation}
admits an extension from $D(0,\varepsilon_0)^\ast \cap \bc^+$
to $D(0,\varepsilon_0)^\ast$, with values in $\sinf \big( \ell^2(\mathcal{V}) 
\big)$ the set of compact linear operators on $\ell^2(\mathcal{V})$. Moreover,
this extension is holomorphic.
\end{lem}

\noindent
\begin{prof}
By Theorem \ref{p4,1}, for $\lambda \in D(0,\varepsilon)^\ast \cap \bc^+$,
$0 < \varepsilon < \frac{\delta}{4}$ small enough, the operator 
\begin{equation}\label{eq5,30}
e_- \big( -L + k + 1 - z_{t_-}(\lambda) \big)^{-1} e_-
\end{equation}
admits the kernel
\begin{equation}\label{eq5,4}
K(\lambda,v,v') = \frac{1}{2 \sqrt{k}} \sqrt{\frac{2}{\pi}} \Big( K_1(\lambda,v,v') 
+ K_2(\lambda,v,v') \Big),
\end{equation}
where 
\begin{equation}\label{eq5,5}
\begin{split}
K_1(\lambda,v,v') : = \sum_{n \, \geq \, 0} \sum_{\substack{j \, \geq \, 0 \\ \ell \, \geq \, 0}} 
\sum_{\substack{0 \leq m \leq N_j \\ 0 \leq q \leq N_\ell}} 
e^{-\frac{\delta}{2}\vert v' \vert} \overline{ E_m^{n,n+j} (v')}
\langle \chi_m^{n,n + j},\chi_q^{n,n + \ell} \rangle 
f_1(j,\ell,\lambda) e^{-\frac{\delta}{2}\vert v \vert} E_q^{n,n + \ell} (v),
\end{split}
\end{equation}
and
\begin{equation}\label{eq5,6}
\begin{split}
K_2(\lambda,v,v') := \sum_{n \, \geq \, 0} \sum_{\substack{j \, \geq \, 0 \\ 
\ell \, \geq \, 0}} \sum_{\substack{0 \leq m \leq N_j \\ 0 \leq q \leq N_\ell}}
e^{-\frac{\delta}{2}\vert v' \vert} \overline{ E_m^{n,n+j} (v')}
\langle \chi_m^{n,n + j},\chi_q^{n,n + \ell} \rangle 
f_2(j,\ell,\lambda) e^{-\frac{\delta}{2}\vert v \vert} E_q^{n,n + \ell} (v),
\end{split}
\end{equation}
with 
\begin{equation}\label{eq5,60}
f_1(j,\ell,\lambda) := - \frac{i e^{i(j + \ell + 2) 2\arcsin \frac{\lambda}{2}}}{\lambda
\sqrt{4 - \lambda^2}} , \qquad 
f_2(j,\ell,\lambda) := \frac{i e^{i\vert j - \ell \vert 2\arcsin \frac{\lambda}{2}}}{\lambda
\sqrt{4 - \lambda^2}}.
\end{equation}

a) We want to prove the convergence of 
$\displaystyle\sum_{v, v' \in \, \mathcal{V}} \big\vert K(\lambda,v,v') \big\vert^2$ for 
$\lambda \in D(0,\varepsilon_0)^\ast$ for some
$0 < \varepsilon_0 \le \frac{\delta}{8}$ small enough.

\smallskip

\noindent
We point out that constants are generic, i.e. can change from an estimate to another. By 
\eqref{eq5,4}--\eqref{eq5,6}, we have
\begin{equation}\label{eq5,7}
\begin{split}
\sum_{v, v' \in \, \mathcal{V}} \big\vert K(\lambda,v,v') \big\vert^2 & = \frac{1}{2k\pi}
\sum_{\substack{r \, \ge \, 0 \\ r' \ge \, 0}}
\sum_{\substack{v \, \in \, S_r \\ v' \in \, S_{r'}}} \big\vert K_1(\lambda,v,v') 
+ K_2(\lambda,v,v') \big\vert^2 \\
& \le \frac{1}{k\pi} \sum_{\substack{r \, \ge \, 0 \\ r' \ge \, 0}}
\sum_{\substack{v \, \in \, S_r \\ v' \in \, S_{r'}}} \left( \big\vert 
K_1(\lambda,v,v') \big\vert^2 + \big\vert K_2(\lambda,v,v') \big\vert^2 \right).
\end{split}
\end{equation}
Let us first prove that $\displaystyle\sum_{\substack{r \, \ge \, 0 \\ r' \ge \, 0}}
\sum_{\substack{v \, \in \, S_r \\ v' \in \, S_{r'}}} \big\vert K_1(\lambda,v,v') \big\vert^2$ 
converges accordingly to the above claim. Thanks to \eqref{eq5,5}, the properties 
(in Section \ref{sec4}) of the vectors $E_m^{n,n+j}$, $\chi_m^{n,n+j}$
and \eqref{eq4,3}, we have 
\begin{equation}\label{eq5,9}
\begin{split}
& \sum_{\substack{r \, \ge \, 0 \\ r' \ge \, 0}}
\sum_{\substack{v \, \in \, S_r \\ v' \in \, S_{r'}}} \big\vert K_1(\lambda,v,v') \big\vert^2 
= \sum_{\substack{r \, \ge \, 0 \\ r' \ge \, 0}}
\sum_{\substack{v \, \in \, S_r \\ v' \in \, S_{r'}}} \Biggl\vert
\sum_{n = 0}^{\min(r,r')} \sum_{\substack{j : n + j = r' \\ \ell : n + \ell = r}} 
\sum_{\substack{0 \leq m \leq N_j \\ 0 \leq q \leq N_\ell}} 
e^{-\frac{\delta}{2} r'} \\
& \times \overline{ E_m^{n,n+j} (v')} \langle \chi_m^{n,n + j},\chi_q^{n,n + \ell} 
\rangle \left( - \frac{i e^{i(j + \ell + 2) 2\arcsin \frac{\lambda}{2}}}{\lambda
\sqrt{4 - \lambda^2}} \right) e^{-\frac{\delta}{2}r} E_q^{n,n + \ell} (v) \Biggr\vert^2 \\
& \le C \sum_{\substack{r \, \ge \, 0 \\ r' \ge \, 0}} 
\sum_{\substack{v \, \in \, S_r \\ v' \in \, S_{r'}}} 
\left( \sum_{n = 0}^{\min(r,r')} \sum_{\substack{j : n + j = r' \\ \ell : n + \ell = r}} 
\sum_{\substack{0 \leq m \leq N_j \\ 0 \leq q \leq N_\ell}} 
e^{-\frac{\delta}{2} r'} \frac{e^{-(j + \ell + 2) \im \big( 2\arcsin \frac{ \lambda}{2}
\big)}}{\big\vert \lambda \sqrt{4 - \lambda^2} \big\vert } e^{-\frac{\delta}{2}r} \right)^2 \\
& \le C \sum_{\substack{r \, \ge \, 0 \\ r' \ge \, 0}}
\sum_{\substack{v \, \in \, S_r \\ v' \in \, S_{r'}}} 
\left( \sum_{n = 0}^{\min(r,r')} \sum_{\substack{j : n + j = r' \\ \ell : n + \ell = r}} 
k^{n + j} k^{n + \ell} e^{-\frac{\delta}{2} r'} 
\frac{e^{-(j + \ell + 2) \im \big( 2\arcsin \frac{\lambda}{2}
\big)}}{\big\vert \lambda \sqrt{4 - \lambda^2} \big\vert} e^{-\frac{\delta}{2}r} \right)^2 \\
& \le C \sum_{\substack{r \, \ge \, 0 \\ r' \ge \, 0}} k^{3r} k^{3r'} e^{-\delta r} 
e^{-\delta r'} \left( \sum_{n = 0}^{\min(r,r')} \sum_{\substack{j : n + j = r' \\ \ell : n + \ell = r}} 
\frac{e^{-(j + \ell + 2) \im \big( 2\arcsin \frac{\lambda}{2}
\big)}}{\big\vert \lambda \sqrt{4 - \lambda^2} \big\vert} \right)^2,
\end{split}
\end{equation}
for some constant $C > 0$. Since for $0 < \vert \lambda \vert \ll 1$ we have 
$2 \arcsin \frac{\lambda}{2} = \lambda + o(\vert \lambda \vert)$, then there exists 
$0 < \varepsilon_0 \le \frac{\delta}{8}$ small enough such that for each 
$0 < \vert \lambda \vert \le \varepsilon_0$, we have
\begin{equation}\label{eq5,90}
\frac{e^{-(j + \ell + 2) \im \big( 2\arcsin \frac{\lambda}{2}
\big)}}{\big\vert \lambda \sqrt{4 - \lambda^2} \big\vert} \le 
\frac{e^{-\big( \im (\lambda) - \frac{\delta}{8} \big) (j + \ell + 2)}}
{\big\vert \lambda \sqrt{4 - \lambda^2} \big\vert}.
\end{equation}
Then, it follows from \eqref{eq5,9} that for each $\lambda \in D(0,\varepsilon_0)^\ast$, 
we have
\begin{equation}\label{eq5,10}
\begin{split}
& \sum_{\substack{r \, \ge \, 0 \\ r' \ge \, 0}} \sum_{\substack{v \, \in \, S_r \\ v' \in \, S_{r'}}} \big\vert K_1(\lambda,v,v') \big\vert^2 \\
& \le C \sum_{\substack{r \, \ge \, 0 \\ r' \ge \, 0}} k^{3r} k^{3r'} 
e^{-\delta r} e^{-\delta r'} 
\left( \sum_{n = 0}^{\min(r,r')} \sum_{\substack{j : n + j = r' \\ \ell : n + \ell = r}}  
\frac{e^{-\big( \im (\lambda) - \frac{\delta}{8} \big) (j + \ell + 2)}}{\big\vert \lambda \sqrt{4 - \lambda^2} \big\vert} \right)^2.
\end{split}
\end{equation}
Clearly, if $F = F(j,\ell)$ is a function of the variables
$j$ and $\ell$, then 
\begin{equation}\label{eq5,11}
\sum_{n = 0}^{\min(r,r')} \sum_{\substack{j : n + j = r' \\ \ell : n + \ell = r}} F(j,\ell)
= \sum_{n = 0}^{\min(r,r')} F(r'-n,r-n).
\end{equation}
This together with \eqref{eq5,10} imply that
\begin{equation}\label{eq5,12}
\begin{split}
& \sum_{\substack{r \, \ge \, 0 \\ r' \ge \, 0}}
\sum_{\substack{v \, \in \, S_r \\ v' \in \, S_{r'}}} \big\vert K_1(\lambda,v,v') \big\vert^2  \\
& \le C \sum_{\substack{r \, \ge \, 0 \\ r' \ge \, 0}} k^{3r} k^{3r'} 
e^{-\delta r} e^{-\delta r'} 
\left( \sum_{n = 0}^{\min(r,r')} \frac{e^{-\big( \im (\lambda) 
- \frac{\delta}{8} \big) (r + r' + 2)} e^{2\big( \im(\lambda) - \frac{\delta}{8} \big)n }} {\big\vert \lambda \sqrt{4 - \lambda^2} \big\vert} 
\right)^2.
\end{split}
\end{equation}
Since $\im (\lambda) - \frac{\delta}{8} \leq 0$, then it follows from \eqref{eq5,12} that
\begin{equation}\label{eq5,13}
\begin{split}
& \sum_{\substack{r \, \ge \, 0 \\ r' \ge \, 0}} 
\sum_{\substack{v \, \in \, S_r \\ v' \in \, S_{r'}}} \big\vert K_1(\lambda,v,v') \big\vert^2 \\
& \le C \sum_{\substack{r \, \ge \, 0 \\ r' \ge \, 0}} k^{3r} k^{3r'} 
e^{-\delta r} e^{-\delta r'} 
\left( \sum_{n = 0}^{\min(r,r')} \frac{e^{-\big( \im (\lambda) 
- \frac{\delta}{8} \big) (r + r' + 2)}}{\big\vert \lambda \sqrt{4 - \lambda^2} \big\vert} \right)^2 \\
& \leq C \sum_{\substack{r \, \ge \, 0 \\ r' \ge \, 0}} k^{3r} k^{3r'} e^{-\delta r} 
e^{-\delta r'} \frac{e^{-2\big( \im (\lambda) - \frac{\delta}{8} \big) (r + r' + 2)}} 
{\big\vert \lambda^2 (4 - \lambda^2) \big\vert} (r + 1)(r' + 1) \\
& = C \sum_{\substack{r \, \ge \, 0 \\ r' \ge \, 0}} (r + 1) k^{3r} e^{-\delta r} 
\frac{e^{-2\big( \im (\lambda) - \frac{\delta}{8} \big) (r + 1)} e^{-2\big( \im (\lambda) - \frac{\delta}{8} \big) (r' + 1)}} {\big\vert \lambda^2 (4 - \lambda^2) \big\vert} 
(r' + 1) k^{3r'} e^{-\delta r'} \\
& \leq C \sum_{\substack{r \, \ge \, 0 \\ r' \ge \, 0}} (r + 1) k^{3(r+1)} e^{-\delta (r+1)} 
\frac{e^{-2\big( \im (\lambda) - \frac{\delta}{8} \big) (r + 1)} e^{-2\big( \im (\lambda) - \frac{\delta}{8} \big) (r' + 1)}} {\big\vert \lambda^2 (4 - \lambda^2) \big\vert} \\
& \hspace{3cm} \times (r' + 1) k^{3(r'+1)} e^{-\delta (r'+1)} \\
& = C \sum_{\substack{r \, \ge \, 0 \\ r' \ge \, 0}} (r + 1) e^{-\big( \frac{\delta}{2} 
- 3 \ln(k) \big) (r+1)} \frac{e^{-2 \big( \im(\lambda) + \frac{\delta}{8} \big) 
(r + 1)} e^{-2\big( \im(\lambda) + \frac{\delta}{8} \big) (r' + 1)}} {\big\vert \lambda^2 (4 - \lambda^2) \big\vert} \\
& \hspace{3cm} \times (r' + 1) e^{-\big( \frac{\delta}{2} - 3 \ln(k) \big) (r'+1)}.
\end{split}
\end{equation}
Assumption (A) implies that $\frac{\delta}{2} - 3 \ln(k) \geq 0$. 
Thus, the r.h.s. and then the l.h.s. of \eqref{eq5,13} is convergent for any
$\lambda \in D(0,\varepsilon_0)^\ast$.

\smallskip

\noindent
Similarly let us prove that $\displaystyle\sum_{\substack{r \, \ge \, 0 \\ r' \ge \, 0}}
\sum_{\substack{v \, \in \, S_r \\ v' \in \, S_{r'}}} \big\vert K_2(\lambda,v,v') \big\vert^2$ 
converges. As in \eqref{eq5,9}, we can show that 
\begin{equation}\label{eq5,14}
\begin{split}
\sum_{\substack{r \, \ge \, 0 \\ r' \ge \, 0}} &
\sum_{\substack{v \, \in \, S_r \\ v' \in \, S_{r'}}}  \big\vert K_2(\lambda,v,v') \big\vert^2 
= \sum_{\substack{r \, \ge \, 0 \\ r' \ge \, 0}}
\sum_{\substack{v \, \in \, S_r \\ v' \in \, S_{r'}}} \Biggl\vert
\sum_{n = 0}^{\min(r,r')} \sum_{\substack{j : n + j = r' \\ \ell : n + \ell = r}} 
\sum_{\substack{0 \leq m \leq N_j \\ 0 \leq q \leq N_\ell}} \\
&  e^{-\frac{\delta}{2} r'} \overline{ E_m^{n,n+j} (v')}
\langle \chi_m^{n,n + j},\chi_q^{n,n + \ell} \rangle 
\left( \frac{i e^{i \vert j - \ell \vert 2\arcsin \frac{\lambda}{2}}}{\lambda
\sqrt{4 - \lambda^2}} \right) e^{-\frac{\delta}{2}r} E_q^{n,n + \ell} (v) \Biggr\vert^2 \\
& \le C \sum_{\substack{r \, \ge \, 0 \\ r' \ge \, 0}} k^{3r} k^{3r'} e^{-\delta r} 
e^{-\delta r'} \left( \sum_{n = 0}^{\min(r,r')} \sum_{\substack{j : n + j = r' \\ \ell 
: n + \ell = r}} \frac{e^{- \vert j - \ell \vert \im \big( 2\arcsin \frac{\lambda}{2}
\big)}}{\big\vert \lambda \sqrt{4 - \lambda^2} \big\vert} \right)^2.
\end{split}
\end{equation}
Thus, similarly to \eqref{eq5,10}, for each $\lambda \in D(0,\varepsilon_0)^\ast$, we have
\begin{equation}\label{eq5,15}
\begin{split}
& \sum_{\substack{r \, \ge \, 0 \\ r' \ge \, 0}} 
\sum_{\substack{v \, \in \, S_r \\ v' \in \, S_{r'}}} \big\vert K_2(\lambda,v,v') \big\vert^2 \\ 
& \le C \sum_{\substack{r \, \ge \, 0 \\ r' \ge \, 0}} k^{3r} k^{3r'} 
e^{-\delta r} e^{-\delta r'} 
\left( \sum_{n = 0}^{\min(r,r')} \sum_{\substack{j : n + j = r' \\ \ell : n + \ell = r}}  \frac{e^{-\big( \im (\lambda) - \frac{\delta}{8} \big) \vert j - \ell \vert}}{\big\vert \lambda \sqrt{4 - \lambda^2} \big\vert} \right)^2.
\end{split}
\end{equation}
In this case, we use \eqref{eq5,11} to write 
\begin{equation}\label{eq5,16}
\begin{split}
& \sum_{\substack{r \, \ge \, 0 \\ r' \ge \, 0}} 
\sum_{\substack{v \, \in \, S_r \\ v' \in \, S_{r'}}} \big\vert K_2(\lambda,v,v') \big\vert^2 \\
& \le C \sum_{\substack{r \, \ge \, 0 \\ r' \ge \, 0}} k^{3r} k^{3r'} 
e^{-\delta r} e^{-\delta r'} 
\left( \sum_{n = 0}^{\min(r,r')} \frac{e^{-\big( \im (\lambda) 
- \frac{\delta}{8} \big) \vert r - r' \vert}} {\big\vert \lambda \sqrt{4 - \lambda^2} \big\vert} \right)^2 \\
& \leq C \sum_{\substack{r \, \ge \, 0 \\ r' \ge \, 0}} k^{3r} k^{3r'} e^{-\delta r} 
e^{-\delta r'} \frac{e^{-2\big( \im (\lambda) - \frac{\delta}{8} \big) \vert r - r' \vert}} 
{\big\vert \lambda^2 (4 - \lambda^2) \big\vert} (r + 1)(r' + 1) \\
& = C \sum_{\substack{r \, \ge \, 0 \\ r' \ge \, 0}} (r + 1) e^{-\big( \frac{\delta}{2} 
- 3 \ln(k) \big)r} \frac{e^{- \frac{\delta}{2}r} e^{-2\big( \im (\lambda) - \frac{\delta}{8} \big) \vert r - r' \vert)} e^{- \frac{\delta}{2}r'}} {\big\vert \lambda^2 (4 - \lambda^2) \big\vert} \\
& \hspace*{3cm} \times (r' + 1) e^{-\big( \frac{\delta}{2} - 3 \ln(k) \big)r'}.
\end{split}
\end{equation}
Assumption (A) implies that $\frac{\delta}{2} - 3 \ln(k) \geq 0$. Thus, the r.h.s. 
and then the l.h.s. of \eqref{eq5,16} is convergent for any
$\lambda \in D(0,\varepsilon_0)^\ast$.

\medskip

\noindent
Now, since $\displaystyle\sum_{v, v' \in \, \mathcal{V}} \big\vert K(\lambda,v,v') \big\vert^2$ 
is convergent for $\lambda \in D(0,\varepsilon_0)^\ast$, 
then the operator given by \eqref{eq5,30} belongs in
$\sd \big( \ell^2(\mathcal{V}) \big)$ for $\lambda \in D(0,\varepsilon_0)^\ast$, the class of Hilbert-Schmidt operators 
on $\ell^2(\mathcal{V})$. Consequently, the operator-valued function defined by 
\eqref{eq5,3} can be extended from $D(0,\varepsilon_0)^\ast \cap \bc^+$ to 
$D(0,\varepsilon_0)^\ast$, with values in $\sinf \big( \ell^2(\mathcal{V}) \big)$. 
It remains to prove that this extension is holomorphic.

\medskip

b) To simply notation, let us denote this extension by 
$$
D(0,\varepsilon_0)^\ast \ni \lambda \mapsto T(\lambda).
$$
Since the kernel of the operator $T(\lambda)$ is given by $K(\lambda,v,v')$ defined
by \eqref{eq5,4}, then to show the claim, it is sufficient to prove it for the
maps 
$$
D(0,\varepsilon_0)^\ast \ni \lambda \mapsto T_s(\lambda),
$$
where for $s = 1, 2$, $T_s(\lambda)$ is the operator with kernel given by 
$K_s(\lambda,v,v')$ in \eqref{eq5,5} and \eqref{eq5,6}. We give the proof only
for the case $s = 1$, the case $s = 2$ being treated in a similar way.
So, for $\lambda \in D(0,\varepsilon_0)^\ast$, let $f_1(j,\ell,\lambda)$ be the
function defined by \eqref{eq5,60} and $D_1(\lambda)$ be the operator whose kernel is
\begin{align*}
& \sum_{n \, \geq \, 0} \sum_{\substack{j \, \geq \, 0 \\ \ell \, \geq \, 0}} 
\sum_{\substack{0 \leq m \leq N_j \\ 0 \leq q \leq N_\ell}} 
e^{-\frac{\delta}{2}\vert v' \vert} \overline{ E_m^{n,n+j} (v')}
\langle \chi_m^{n,n + j},\chi_q^{n,n + \ell} \rangle \partial_\lambda f_1(j,\ell,\lambda)
e^{-\frac{\delta}{2}\vert v \vert} E_q^{n,n + \ell} (v) \\
& = \sum_{n \, \geq \, 0} \sum_{\substack{j \, \geq \, 0 \\ \ell \, \geq \, 0}} 
\sum_{\substack{0 \leq m \leq N_j \\ 0 \leq q \leq N_\ell}} 
e^{-\frac{\delta}{2}\vert v' \vert} \overline{ E_m^{n,n+j} (v')}
\langle \chi_m^{n,n + j},\chi_q^{n,n + \ell} \rangle \\
& \times \left( - \frac{i e^{i(j + \ell + 2) 2\arcsin \frac{\lambda}{2}}}{\lambda^2
(4 - \lambda^2)} \right) e^{-\frac{\delta}{2}\vert v \vert} E_q^{n,n + \ell} (v)
\Bigg( 2i (j + \ell + 2) - \frac{4 - 2\lambda^2}{\sqrt{4 - \lambda^2}} \Bigg).
\end{align*}
As in a) above, we can show that $D_1(\lambda) \in \sd \big( \ell^2(\mathcal{V}) \big)$.
Therefore, for $\lambda_0 \in D(0,\varepsilon_0)^\ast$, the kernel of the
Hilbert-Schmidt operator $\frac{T_1(\lambda) - T_1(\lambda_0)}{\lambda - \lambda_0} 
- D_1(\lambda_0)$ is given by
\begin{equation}\label{eq5,160}
\begin{split}
& K_1(\lambda,\lambda_0,v,v') := \sum_{n \, \geq \, 0} \sum_{\substack{j \, 
\geq \, 0 \\ \ell \, \geq \, 0}} 
\sum_{\substack{0 \leq m \leq N_j \\ 0 \leq q \leq N_\ell}} 
e^{-\frac{\delta}{2}\vert v' \vert} \overline{ E_m^{n,n+j} (v')}
\langle \chi_m^{n,n + j},\chi_q^{n,n + \ell} \rangle \\
& \times \Bigg( \frac{f_1(j,\ell,\lambda) - f_1(j,\ell,\lambda_0)}{\lambda - \lambda_0} - 
\partial_\lambda f_1(j,\ell,\lambda_0) \Bigg) e^{-\frac{\delta}{2}\vert v \vert} E_q^{n,n + \ell} (v).
\end{split}
\end{equation}
Thus, to conclude the proof of the lemma, we have just to justify that
$$
\Bigg\Vert \frac{T_1(\lambda) - T_1(\lambda_0)}{\lambda - \lambda_0} 
- D_1(\lambda_0) \Bigg\Vert_{\sd(\ell^2(\mathcal{V}))} \longrightarrow 0
$$ 
as $\lambda \rightarrow \lambda_0$. Since we have
\begin{equation}\label{eq5,161}
\bigg\Vert \frac{T_1(\lambda) - T_1(\lambda_0)}{\lambda - \lambda_0} 
- D_1(\lambda_0) \bigg\Vert_{\sd(\ell^2(\mathcal{V}))} \le 
\sum_{v, v' \in \, \mathcal{V}} \big\vert K_1(\lambda,\lambda_0,v,v') \big\vert^2,
\end{equation}
then it suffices to prove that the r.h.s. of \eqref{eq5,161} tends to zero
as $\lambda \rightarrow \lambda_0$. The Taylor-Lagrange formula applied to the 
function 
$$
[0,1] \ni t \mapsto g(t) := f_1 \big(j,\ell,t\lambda + (1-t)\lambda_0 \big) 
$$
asserts there exists $\theta \in (0,1)$ such that
\begin{equation}\label{eq5,162}
\begin{split}
f_1(j,\ell,\lambda) = f_1(j,\ell,\lambda_0) + (\lambda - \lambda_0) 
\partial_\lambda f_1(j,\ell,\lambda_0) + \frac{(\lambda - \lambda_0)^2}{2} 
\partial_\lambda^{(2)} f_1 \big(j,\ell, \theta \lambda + (1-\theta)\lambda_0 \big).
\end{split}
\end{equation}
Then, it follows from \eqref{eq5,160} and \eqref{eq5,162} that $K_1(\lambda,\lambda_0,v,v')$ 
can be represented as
\begin{equation}\label{eq5,163}
\begin{split}
& K_1(\lambda,\lambda_0,v,v') = \frac{\lambda - \lambda_0}{2} 
\sum_{n \, \geq \, 0} \sum_{\substack{j \, 
\geq \, 0 \\ \ell \, \geq \, 0}} 
\sum_{\substack{0 \leq m \leq N_j \\ 0 \leq q \leq N_\ell}} 
e^{-\frac{\delta}{2}\vert v' \vert} \overline{ E_m^{n,n+j} (v')}
\langle \chi_m^{n,n + j},\chi_q^{n,n + \ell} \rangle \\
& \times \partial_\lambda^{(2)} f_1 \big(j,\ell, \theta \lambda + (1-\theta)\lambda_0 \big) e^{-\frac{\delta}{2}\vert v \vert} E_q^{n,n + \ell} (v).
\end{split}
\end{equation}
Now, easy but fastidious computations allow to see that there exists a family 
of holomorphic functions $F_{p,q}$, $0 \le p \le q$, on $D(0,\varepsilon_0)^\ast$ such that
$$
\partial_\lambda^{(q)} f_1(j,\ell,\lambda) = ie^{i(j+\ell+2) 2\arcsin \frac{\lambda}{2}}
\sum_{p=0}^q F_{p,q}(\lambda) (j + \ell + 2)^{q-p}.
$$
In particular, for $q = 2$, we have
$$
\partial_\lambda^{(2)} f_1(j,\ell,\lambda) = ie^{i(j+\ell+2) 2\arcsin \frac{\lambda}{2}}
\Big( F_{0,2}(\lambda)(j+\ell+2)^2 + F_{1,2}(\lambda) (j+\ell+2) + F_{2,2}(\lambda) \Big).
$$
Putting this together with \eqref{eq5,163}, we obtain
\begin{align*}
& \big\vert K_1(\lambda,\lambda_0,v,v') \big\vert \\
& \le \frac{\vert \lambda - \lambda_0 \vert}{2} \big\vert F_{0,2} \big( \theta \lambda + (1-\theta)\lambda_0 \big) \big\vert \sum_{n \, \geq \, 0} \sum_{\substack{j \, 
\geq \, 0 \\ \ell \, \geq \, 0}} 
\sum_{\substack{0 \leq m \leq N_j \\ 0 \leq q \leq N_\ell}} 
\bigg\vert e^{-\frac{\delta}{2}\vert v' \vert} \overline{ E_m^{n,n+j} (v')}
\langle \chi_m^{n,n + j},\chi_q^{n,n + \ell} \rangle \\
& \hspace*{2cm} \times e^{i(j+\ell+2) 2\arcsin \frac{\theta \lambda + (1-\theta)\lambda_0}{2}}
(j+\ell+2)^2 e^{-\frac{\delta}{2}\vert v \vert} E_q^{n,n + \ell} (v) \bigg\vert \\
& + \frac{\vert \lambda - \lambda_0 \vert}{2} \big\vert F_{1,2} \big( \theta \lambda + (1-\theta)\lambda_0 \big) \big\vert \Big\vert \sum_{n \, \geq \, 0} \sum_{\substack{j \, 
\geq \, 0 \\ \ell \, \geq \, 0}} 
\sum_{\substack{0 \leq m \leq N_j \\ 0 \leq q \leq N_\ell}} 
\bigg\vert e^{-\frac{\delta}{2}\vert v' \vert} \overline{ E_m^{n,n+j} (v')}
\langle \chi_m^{n,n + j},\chi_q^{n,n + \ell} \rangle \\
& \hspace*{2cm} \times e^{i(j+\ell+2) 2\arcsin \frac{\theta \lambda + (1-\theta)\lambda_0}{2}}
(j+\ell+2) e^{-\frac{\delta}{2}\vert v \vert} E_q^{n,n + \ell} (v) \bigg\vert \\
& + \frac{\vert \lambda - \lambda_0 \vert}{2} \big\vert F_{2,2} \big( \theta \lambda + (1-\theta)\lambda_0 \big) \big\vert \sum_{n \, \geq \, 0} \sum_{\substack{j \, 
\geq \, 0 \\ \ell \, \geq \, 0}} 
\sum_{\substack{0 \leq m \leq N_j \\ 0 \leq q \leq N_\ell}} 
\bigg\vert e^{-\frac{\delta}{2}\vert v' \vert} \overline{ E_m^{n,n+j} (v')}
\langle \chi_m^{n,n + j},\chi_q^{n,n + \ell} \rangle \\
& \hspace*{2cm} \times e^{i(j+\ell+2) 2\arcsin \frac{\theta \lambda + (1-\theta)\lambda_0}{2}}
 e^{-\frac{\delta}{2}\vert v \vert} E_q^{n,n + \ell} (v) \bigg\vert \\
& =: \sum_{p=0}^2 Q_p(\lambda,\lambda_0,v,v').
\end{align*}
Thus, we have
\begin{equation}\label{eq5,164}
\begin{split}
& \sum_{v, v' \in \, \mathcal{V}} \big\vert K_1(\lambda,\lambda_0,v,v') \big\vert^2 \\
& \le {\rm Const.} \Bigg( \sum_{v, v' \in \, \mathcal{V}} Q_1(\lambda,\lambda_0,v,v')^2
+ \sum_{v, v' \in \, \mathcal{V}} Q_2(\lambda,\lambda_0,v,v')^2 + 
\sum_{v, v' \in \, \mathcal{V}} Q_3(\lambda,\lambda_0,v,v')^2 \Bigg).
\end{split}
\end{equation}
For $p \in \lbrace 0,1,2 \rbrace$, let us show that $\displaystyle\sum_{v, v' \in \, \mathcal{V}} Q_p(\lambda,\lambda_0,v,v')^2 \longrightarrow 0$ as $\lambda \rightarrow \lambda_0$. 
Since  $\theta \lambda + (1-\theta)\lambda_0$ belongs in $ D(0,\varepsilon_0)^\ast$ for 
$\lambda, \lambda_0 \in D(0,\varepsilon_0)^\ast$, then similarly to \eqref{eq5,90} we have
\begin{equation}\label{eq5,165}
\begin{split}
e^{-(j + \ell + 2) \im \big( 2\arcsin \frac{\theta \lambda + (1-\theta)\lambda_0}{2}
\big)} & \le e^{-\big( \im (\theta \lambda + (1-\theta)\lambda_0) - \frac{\delta}{8} \big) (j + \ell + 2)} \\
& \le e^{\frac{\delta}{4}(j + \ell + 2)}.
\end{split}
\end{equation}
Therefore, by arguing as in a) above, we obtain that there exists a uniform 
constant Const. in $\lambda$ and $\lambda_0$ such that for $q \in \lbrace 0,1,2 \rbrace$,
$$
\sum_{v, v' \in \, \mathcal{V}} Q_p(\lambda,\lambda_0,v,v')^2 \le {\rm Const.} 
\frac{\vert \lambda - \lambda_0 \vert^2}{4} \big\vert F_{q,2}(\theta \lambda + (1-\theta)\lambda_0) \big\vert^2 \underset{\lambda \rightarrow \lambda_0}{\longrightarrow} 0,
$$
implying by \eqref{eq5,161} and \eqref{eq5,164} that $\Big\Vert \frac{T_1(\lambda) - T_1(\lambda_0)}{\lambda - \lambda_0} - D_1(\lambda_0) \Big\Vert_{\sd(\ell^2(\mathcal{V}))} \longrightarrow 0$ as $\lambda \rightarrow \lambda_0$. Thus, the operator-valued function 
$D(0,\varepsilon_0)^\ast \ni \lambda \mapsto T_1(\lambda)$ is holomorphic with derivative 
$\partial_\lambda T_1(\lambda) = D_1(\lambda)$. Similarly, 
$D(0,\varepsilon_0)^\ast \ni \lambda \mapsto T_2(\lambda)$ is holomorphic, and then $D(0,\varepsilon_0)^\ast \ni \lambda \mapsto T(\lambda)$. This concludes the proof of the lemma.
\end{prof}

It follows from the identities
\begin{equation}\label{eq5,17}
\Big( \ch_{\pm \widetilde M} - z \Big)^{-1} \left( I \pm {\widetilde M} (-L + k + 1 - z)^{-1} 
\right) = (-L + k + 1 - z)^{-1},
\end{equation} 
that
\begin{equation}\label{eq5,18}
\begin{split}
e_- \Big( & \ch_{\pm \widetilde M} - z \Big)^{-1} e_- \\
& = e_- (-L + k + 1 - z)^{-1} e_- \left( I \pm e_+ {\widetilde M}(-L + k + 1 - z)^{-1} 
e_- \right)^{-1}.
\end{split}
\end{equation}
Assumption (A) on the potential perturbation $M$ implies that 
\begin{equation}\label{eq5,19}
e_+ {\widetilde M} = \mathscr{\widetilde M} e_-
\end{equation}
for some bounded operator $\mathscr{\widetilde M}$ on $\ell^2(\mathcal{V})$. Thus, combining \eqref{eq5,19} and Lemma \ref{l5,1}, we obtain that the operator-valued functions
\begin{equation}\label{eq5,20}
\lambda \longmapsto \pm e_+ {\widetilde M} \big( -L + k + 1 - z_{t_-}(\lambda) \big)^{-1} e_-
\end{equation}
are holomorphic in $D(0,\varepsilon)^\ast$, with values in $\sinf \big( \ell^2(\mathcal{V}) 
\big)$. Therefore, by the analytic Fredholm extension theorem, the operator-valued functions
\begin{equation}\label{eq5,21}
\lambda \longmapsto \left( I \pm e_+ {\widetilde M} \big( -L + k + 1 - z_{t_-}(\lambda) 
\big)^{-1} e_- \right)^{-1}
\end{equation}
admit meromorphic extensions from $D(0,\varepsilon_0)^\ast \cap \bc^+$ to 
$D(0,\varepsilon_0)^\ast$. Defining the Banach spaces 
\begin{equation}\label{eq5,22}
\ell_{\pm \delta}^{2}(\mathcal{V}) := e^{\pm \frac{\delta}{2} \vert v \vert} \ell^2(\mathcal{V}),
\end{equation}
we then get the following proposition:
\begin{prop}\label{p5,1} 
The operator-valued functions
\begin{equation}\label{eq5,23}
\lambda \longmapsto \Big( \ch_{\pm \widetilde M} - z_{t_-}(\lambda) \Big)^{-1} \in  
\mathscr{L} \left( \ell_{-\delta}^{2}(\mathcal{V}),\ell_\delta^{2}(\mathcal{V}) \right),
\end{equation} 
admit meromorphic extensions from $D(0,\varepsilon_0)^\ast \cap \bc^+$ to 
$D(0,\varepsilon_0)^\ast$. These extensions will be denoted by $R_{\pm \widetilde M} 
\big( z_{t_-}(\lambda) \big)$ respectively.
\end{prop}

\noindent
As in \eqref{eq5,19}, Assumption (A) on $M$ implies that there exists a bounded operator 
$\mathscr{B}$ on $\ell^2(\mathcal{V})$ such that $\sqrt{\vert \widetilde M \vert} = \mathscr{B} e_-$.
Together with Lemma \ref{l5,1}, this gives the following lemma:

\begin{lem}\label{l5,2} 
Let $J$ be defined by the polar decomposition ${\widetilde M} = J \vert \widetilde M \vert$ 
of the potential perturbation ${\widetilde M}$. Then, the operator-valued functions
\begin{equation}\label{eq5,24}
\lambda \longmapsto \mathcal{T}_{\pm \widetilde M} \big( z_{t_-}(\lambda) \big) := \pm J 
\sqrt{\vert \widetilde M \vert} \big( -L + k + 1 - z_{t_-}(\lambda) \big)^{-1} 
\sqrt{\vert \widetilde M \vert},
\end{equation} 
admit holomorphic extensions from $D(0,\varepsilon_0)^\ast \cap \bc^+$
to $D(0,\varepsilon_0)^\ast$, with values in $\sinf \big( \ell^2(\mathcal{V}) \big)$.
\end{lem}

\noindent
We are now in position to define the resonances of the operator $\ch_{\widetilde M}$ near 
the spectral thresholds $z = t_{\pm}(k)$. Note that in the next definitions, the quantity 
$Ind_{\gamma}(\cdot)$ is defined in the appendix by \eqref{eqa,2}.

\begin{de}\label{d5,1} 
We define the resonances of the operator $\ch_{\widetilde M}$ near $t_-(k)$ as the poles 
of the meromorphic extension $R_{\widetilde M}(z)$, of the resolvent 
$\Big( \ch_{\widetilde M} - z \Big)^{-1}$ in $\mathscr{L} \left( \ell_{-\delta}^{2}(\mathcal{V}),\ell_{\delta}^{2}(\mathcal{V}) \right)$. 
The multiplicity of a resonance 
$$
z_{t_-} := z_{t_-}(\lambda) = t_-(k) + \lambda^2\sqrt{k},
$$ 
is defined by 
\begin{equation}\label{eq5,25}
\textup{mult} \big( z_{t_-} \big) := Ind_{\gamma} \, \biggl( I + 
\mathcal{T}_{\widetilde M} \Big( z_{t_-}(\cdot) \Big) \biggr),
\end{equation}
where $\gamma$ is a small contour positively oriented containing $\lambda$ as the only 
point satisfying that $z_{t_-}(\lambda)$ is a resonance of $\ch_{\widetilde M}$.
\end{de}

As mentioned previously, to define the resonances of the operator $\ch_{\widetilde M}$ near 
$t_+(k)$, there exists a specific reduction which exploits a simple relation between the 
two thresholds $t_{\pm}(k)$. Indeed, define the self-adjoint unitary operator $\Theta$ 
on $\ell^2(\mathcal{V})$ by 
\begin{equation}\label{eq5,26}
(\Theta \varphi)(v) := (-1)^{\vert v \vert} \varphi(v).
\end{equation}
We thus have
\begin{itemize}
\item $\Theta^2 = I$,
\item $\Theta L \Theta^{-1} = -L$,
\item $\Theta \widetilde{M} \Theta^{-1} = \widetilde{M}$.
\end{itemize}
In the last point, we have used the fact that $\widetilde{M}$ is the multiplication operator 
by the function $\widetilde{M}$. Thus, it can be easily verified that we have
\begin{equation}\label{eq5,27}
\Theta \Big( -L + k + 1 + \widetilde{M} - z \Big) \Theta^{-1} = 
-(-L + k + 1) + \widetilde{M} + 2(k + 1) - z,
\end{equation}
so that
\begin{equation}\label{eq5,28}
\Theta e_- \Big( \ch_{\widetilde M} - z \Big)^{-1} e_- \Theta^{-1}
= - e_- \Big( \ch_{-\widetilde M} - \big( 2(k + 1) - z \big) \Big)^{-1} e_-.
\end{equation}
Set 
\begin{equation}\label{eq5,29}
\omega := 2(k + 1) - z. 
\end{equation}
Since $\omega$ is near $t_-(k)$ for $z$ near $t_+(k)$, then using relation \eqref{eq5,28}, 
we can define the resonances of the operator $\ch_{\widetilde M}$ near $t_+(k)$ as 
the poles of the meromorphic extension of the resolvent
\begin{equation}\label{eq5,300}
- \Big( \ch_{-\widetilde M} - \omega \Big)^{-1} : \ell_{-\delta}^2(\mathcal{V}) 
\rightarrow \ell_{\delta}^2(\mathcal{V}),
\end{equation}
near $\omega = t_-(k)$, similarly to Definition \ref{d5,1}. More precisely, we have the 
following definition:

\begin{de}\label{d5,2} 
We define the resonances of the operator $\ch_{\widetilde M}$ near $t_+(k)$ as the poles 
of the meromorphic extension $R_{-\widetilde M}(\omega)$, of the resolvent 
$\Big( \ch_{-\widetilde M} - \omega \Big)^{-1}$ in $\mathscr{L} \left( \ell_{-\delta}^{2}(\mathcal{V}),\ell_{\delta}^{2}(\mathcal{V}) \right)$, for $\omega$ given by \eqref{eq5,29} near $t_-(k)$. 
The multiplicity of a resonance 
$$
z_{t_+} := z_{t_+}(\lambda) = 2(k+ 1) - \Big( t_-(k) + \lambda^2\sqrt{k} \Big) 
= t_+(k) - \lambda^2\sqrt{k},
$$ 
is defined by 
\begin{equation}\label{eq5,31}
\textup{mult}(z_{t_+}) := Ind_{\gamma} \, \biggl( I + 
\mathcal{T}_{-\widetilde M} \Big( 2(k + 1) - z_{t_+}(\cdot) \Big) \biggr),
\end{equation}
where $\gamma$ is a small contour positively oriented containing $\lambda$ as the only 
point satisfying that $2(k + 1) - z_{t_+}(\lambda)$ is a pole of $R_{-\widetilde M}(\omega)$.
\end{de}

\begin{rem} 
Notice that the resonances $z_{t_\pm}(\lambda)$ near the spectral thresholds $t_\pm(k)$ 
are defined in some two-sheets Riemann surfaces $\mathcal{M}_{t_\pm}$ respectively.
Otherwise, the discrete eigenvalues of the operator $\ch_{\widetilde M}$ near $t_\pm(k)$ 
are resonances. Moreover, the algebraic multiplicity \eqref{eq1,9} of a discrete eigenvalue 
coincides with its multiplicity as a resonance near $t_\pm(k)$ respectively given by 
\eqref{eq5,25} and \eqref{eq5,31}. Let us give the proof only for the equality 
$\eqref{eq1,9} = \eqref{eq5,25}$, the equality $\eqref{eq1,9} = \eqref{eq5,31}$ could be 
treated in a similar fashion. Let $z_{t_-} := z_{t_-}(\lambda) \in \bc \setminus 
[t_-(k),t_+(k)]$ be a discrete eigenvalue of $\ch_{\widetilde M}$ near $t_-(k)$. 
Firstly, observe that Assumption (A) on $M$ implies that ${\widetilde M}$ is of 
trace-class. In this case, it is is well know (see e.g. \cite[Chap. 9]{si}) that
$z_{t_-} \in \sigma_{\textup{\textbf{disc}}} \big( \ch_{\widetilde M} \big)$ if and 
only if $h(z_{t_-}) = 0$, where for $z \in \bc \setminus [t_-(k),t_+(k)]$, $h$ is the 
holomorphic function defined by
\begin{align*}
h(z) := \textup{det} \Bigl( I + {\widetilde M} \big( -L + k + 1 - z \big)^{-1} \Bigr) 
= \det \Bigg( I + J \sqrt{\vert \widetilde M \vert} \big( -L + k + 1 - z \big)^{-1} 
\sqrt{\vert \widetilde M \vert} \Bigg).
\end{align*}
Moreover, the algebraic multiplicity \eqref{eq1,9} of $z_{t_-}$ is equal to its 
order as zero of the function $h$. Namely, by the residues theorem,
$$
\textup{m}(z_{t_-}) = ind_{\gamma'} h := 
\frac{1}{2i\pi} \int_{\gamma'} \frac{h'(z)}{h(z)} dz,
$$
where $\gamma'$ is a small circle positively oriented containing $z_{t_-}$ as the only 
zero of $h$. Then, the claim follows directly from the equality 
$$
ind_{\gamma'} h = Ind_{\gamma} \, \biggl( I + \mathcal{T}_{\widetilde M} \Big( z_{t_-}(\cdot) \Big) \biggr),
$$
see for instance \cite[Identity (6)]{bbr2} for more details. 
\end{rem}

\subsection{Characterization of the resonances}

In this subsection, we give a simple characterization of resonances of 
$\ch_{\widetilde M}$ near the spectral thresholds $t_\pm(k)$. 
The first one concerns the resonances near $z = t_-(k)$.

\begin{prop}\label{p5,2}
The following assertions are equivalent:

\begin{itemize}
\item[(a)] $z_{t_-} = z_{t_-}(\lambda) \in \mathcal{M}_{t_-}$ is a resonance,
\item[(b)] $z_{t_-}$ is a pole of $R_{\widetilde M}(z)$, 
\item[(c)] $-1$ is an eigenvalue of $\mathcal{T}_{\widetilde M} \big( z_{t_-}(\lambda) \big)$.
\end{itemize}
\end{prop}

\noindent
\begin{prof}
$(a) \Longleftrightarrow (b)$ is just the Definition \ref{d5,1}, while 
$(b) \Longleftrightarrow (c)$ is a consequence of the identity
\begin{equation}
\left( I + J \sqrt{\vert \widetilde M \vert} (-L + k + 1 - z)^{-1} 
\sqrt{\vert \widetilde M \vert} \right) 
\left( I - J \sqrt{\vert \widetilde M \vert} \Big( \ch_{\widetilde M} - z \Big)^{-1} \sqrt{\vert \widetilde M \vert} \right) = I,
\end{equation}
coming from the resolvent equation.
\end{prof}

\noindent
Similarly, we have the following proposition:

\begin{prop}\label{p5,3}
The following assertions are equivalent:

\begin{itemize}
\item[(a)] $z_{t_+} = z_{t_+}(\lambda) \in \mathcal{M}_{t_+}$ is a resonance,
\item[(b)] $2(k + 1) - z_{t_+}(\lambda)$ is a pole of $R_{-\widetilde M}(\omega)$ 
for $\omega$ given by \eqref{eq5,29} near $t_-(\lambda)$,
\item[(c)] $-1$ is an eigenvalue of 
$\mathcal{T}_{-\widetilde M} \Big( 2(k + 1) - z_{t_+}(\lambda) \Big)$.
\end{itemize}
\end{prop}

\section{Proof of Theorem \ref{t1}}\label{sec6}

This section is devoted to the proof of Theorem \ref{t1}. It will be divided into 
tree steps.

\subsection{A preliminary result}\label{ss6,1}
The first step consists on refining the representations of the sandwiched resolvents
$\mathcal{T}_{\widetilde M} \big( z_{t_-}(\lambda) \big)$ and 
$\mathcal{T}_{-\widetilde M} \big( 2(k + 1) - z_{t_+}(\lambda) \big)$
near the spectral thresholds $z = t_{\pm}(k)$. 
Notice that
\begin{equation}\label{eq6,0}
2(k + 1) - z_{t_+}(\lambda) = z_{t_-}(\lambda),
\end{equation} 
so that our analysis will be just reduced to the operators 
$\mathcal{T}_{\pm \widetilde M} \big( z_{t_-}(\lambda) \big)$. 

\medskip

\noindent
Recall that $\mathcal{T}_{\pm \widetilde M} \big( z_{t_-}(\lambda) \big) = \pm J 
\sqrt{\vert \widetilde M \vert} \big( -L + k + 1 - z_{t_-}(\lambda) \big)^{-1} 
\sqrt{\vert \widetilde M \vert}$, and let us set
\begin{equation}\label{eq6,1}
\gamma(\lambda) := \frac{\left( e^{i (j + \ell + 2) 2\arcsin \frac{\lambda}{2}} 
- 1 \right)} {\lambda \sqrt{4 - \lambda^2}} \quad \text{and} \quad
\beta(\lambda) := \frac{\left( e^{i \vert j - \ell \vert 
2\arcsin \frac{\lambda}{2}} - 1 \right)} {\lambda \sqrt{4 - \lambda^2}}, \, 
j, \ell \ge 0.
\end{equation}
By construction, as shows the proof of Lemma \ref{l5,1}, for 
$\lambda \in D(0,\varepsilon_0)^\ast$ the operator 
\begin{equation}\label{eq6,2}
\sqrt{\vert \widetilde M \vert} \big( -L + k + 1 - z_{t_-}(\lambda) \big)^{-1} 
\sqrt{\vert \widetilde M \vert}
\end{equation}
admits the integral kernel 
\begin{equation}\label{eq6,3}
\pm \frac{1}{2 \sqrt{k}} \sqrt{\frac{2}{\pi}} 
\Big( \mathcal{K}_1^{(\lambda)}(v,v') + \mathcal{K}_2^{(\lambda)}(v,v') \Big),
\end{equation}
where 
\begin{equation}\label{eq6,4}
\begin{split}
\mathcal{K}_1^{(\lambda)}(v,v') : = & \sum_{n \, \geq \, 0} 
\sum_{\substack{j \, \geq \, 0 \\ \ell \, \geq \, 0}} 
\sum_{\substack{0 \leq m \leq N_j \\ 0 \leq q \leq N_\ell}} 
\sqrt{\vert \widetilde M \vert} (v') \overline{ E_m^{n,n+j} (v')}
\langle \chi_m^{n,n + j},\chi_q^{n,n + \ell} \rangle \\
& \times i \left( - \gamma(\lambda) - \frac{1}{\lambda \sqrt{4 - \lambda^2}} \right) \sqrt{\vert \widetilde M \vert} (v) E_q^{n,n + \ell} (v),
\end{split}
\end{equation}
and
\begin{equation}\label{eq6,5}
\begin{split}
\mathcal{K}_2^{(\lambda)}(v,v') : = & \sum_{n \, \geq \, 0} 
\sum_{\substack{j \, \geq \, 0 \\ \ell \, \geq \, 0}} 
\sum_{\substack{0 \leq m \leq N_j \\ 0 \leq q \leq N_\ell}}
\sqrt{\vert \widetilde M \vert} (v') \overline{ E_m^{n,n+j} (v')}
\langle \chi_m^{n,n + j},\chi_q^{n,n + \ell} \rangle \\
& \times i \left( \beta(\lambda) + \frac{1}{\lambda \sqrt{4 - \lambda^2}} \right) \sqrt{\vert \widetilde M \vert} (v) E_q^{n,n + \ell} (v).
\end{split}
\end{equation}
Since $\gamma$ and $\beta$ can be extended to holomorphic functions on the open disk 
$D(0,\varepsilon_0)^\ast \cup \lbrace 0 \rbrace$, then by combining identities 
\eqref{eq6,2}-\eqref{eq6,5}, we get the following result:

\begin{prop}\label{p6,1} 
For $\lambda \in D(0,\varepsilon_0)^\ast \cup \lbrace 0 \rbrace$, we have
\begin{equation}\label{eq6,6}
\mathcal{T}_{\pm \widetilde M} \big( z_{t_-}(\lambda) \big) = 
\pm J \sqrt{\frac{2}{\pi}} \, {\rm Hol}(\lambda), 
\end{equation}
where ${\rm Hol}(\lambda)$ defines a holomorphic operator on 
$D(0,\varepsilon_0)^\ast \cup \lbrace 0 \rbrace$ with values in 
$\sinf \big( \ell^2(\mathcal{V}) \big)$, and with kernel given by
\begin{equation}\label{eq6,7}
\begin{split}
& \frac{i}{2 \sqrt{k}} \sum_{n \, \geq \, 0} 
\sum_{\substack{j \, \geq \, 0 \\ \ell \, \geq \, 0}} 
\sum_{\substack{0 \leq m \leq N_j \\ 0 \leq q \leq N_\ell}}
\sqrt{\vert \widetilde M \vert} (v') \overline{ E_m^{n,n+j} (v')}
\langle \chi_m^{n,n + j},\chi_q^{n,n + \ell} \rangle \\
& \hspace*{1cm} \times \Big( \beta(\lambda) - \gamma(\lambda) \Big) \sqrt{\vert \widetilde M \vert} (v) E_q^{n,n + \ell} (v).
\end{split}
\end{equation}
\end{prop}

\subsection{Reformulation of the problem}

Let $\mathscr{H}$ be a separable Hilbert space, $\mathcal{D} \subseteq \bc$ be a 
domain containing $0$, and $\sinf(\mathscr{H})$ denote the set of compact 
linear operators in $\mathscr{H}$. For a holomorphic operator-valued function 
\begin{equation}
K : \mathcal{D} \setminus \lbrace 0 \rbrace \longrightarrow \sinf(\mathscr{H}),
\end{equation}
and a subset $\Omega \subseteq \mathcal{D} \setminus \lbrace 0 \rbrace$, a complex 
number $\lambda \in \Omega$ is said to be a {\em characteristic value} of the 
operator-valued function
\begin{equation} 
\lambda \longmapsto I + K(\lambda),
\end{equation} 
if the operator $I + K(\lambda)$ is not invertible (cf. Section \ref{sa} for more 
details about the concept of characteristic value). By abuse of language,
we shall sometimes say that $\lambda$ is a characteristic value of the operator $I + K(\lambda)$. 
Once there exists $\lambda_{0} \in \Omega$ such that $I + K(\lambda_0)$ is invertible, 
then by the analytic Fredholm theorem, the set of characteristic values 
$\lambda \in \Omega$ of $I + K(\cdot)$ is discrete. Moreover, according to Definition
\ref{d,a1} and \eqref{eqa,2}, the multiplicity of a characteristic value $\lambda$ is 
defined by
\begin{equation}\label{eqa,11}
\textup{mult}(\lambda) := Ind_{\gamma} \big( I + K(\cdot) \big),
\end{equation}
$\gamma$ being a small contour positively oriented which contains $\lambda$ as the only 
point satisfying $I + K(z)$ is not invertible, and with $I + K(\cdot)$ not vanishing 
on $\gamma$. We then can reformulate Propositions \ref{p5,2} and \ref{p5,3} in the
following way:

\begin{prop}\label{p6,2}
For $\lambda \in D(0,\varepsilon_0)^\ast$, the following assertions are 
equivalent:

\begin{itemize}
\item[(a)] $z_{t_-} = z_{t_-}(\lambda) \in \mathcal{M}_{t_-}$ is a resonance,
\item[(b)] $\lambda$ is a characteristic value of  
$I + \mathcal{T}_{\widetilde M} \big( z_{t_-}(\cdot) \big)$. 

\noindent
Moreover, thanks to \eqref{eq5,25}, the multiplicity of the resonance $z_{t_-}(\lambda)$ 
coincides with that of the characteristic value $\lambda$.
\end{itemize}
\end{prop}

\begin{prop}\label{p6,3}
For $\lambda \in D(0,\varepsilon_0)^\ast$, the following assertions are 
equivalent:

\begin{itemize}
\item[(a)] $z_{t_+} = z_{t_+}(\lambda) \in \mathcal{M}_{t_+}$ is a resonance,
\item[(b)] $\lambda$ is a characteristic value of 
$I + \mathcal{T}_{-\widetilde M} \Big( 2(k + 1) - z_{t_+}(\cdot) \Big)$.

\noindent
Moreover, thanks to \eqref{eq5,31}, the multiplicity of the resonance $z_{t_+}(\lambda)$ 
coincides with that of the characteristic value $\lambda$.
\end{itemize}
\end{prop}

\subsection{End of the proof of Theorem \ref{t1}}

From Propositions \ref{p6,2}, \ref{p6,3} and \ref{p6,1} together with the identity 
\eqref{eq6,0}, it follows that $z_{t_\pm}(\lambda)$ is a resonance of $\ch_{\widetilde M}$ 
near $t_\pm(k)$ if and only if $\lambda$ is a characteristic value of 
\begin{equation}
I + \mathcal{T}_{\pm \widetilde M} \big( z_{t_-}(\lambda) \big) = I \pm 
J \sqrt{\frac{2}{\pi}} \, {\rm Hol}(\lambda).
\end{equation}
Since the operator ${\rm Hol}(\lambda)$ is holomorphic in the open disk 
$D(0,\varepsilon_0)^\ast \cup \lbrace 0 \rbrace$ with values in $\sinf \big( \ell^2(\mathcal{V}) 
\big)$, then Theorem \ref{t1} holds by applying Proposition \ref{p,a1} with 
\begin{itemize}
\item $\mathcal{D} = \Omega_r^\ast \cup \lbrace 0 \rbrace$, $\Omega_r^\ast \subseteq D(0,\varepsilon_0)^\ast$,
\item $Z = \lbrace 0 \rbrace$,
\item $F = I + \mathcal{T}_{\pm \widetilde M} \big( z_{t_-}(\cdot) \big)$.
\end{itemize}
This concludes the proof of Theorem \ref{t1}.


\section{Appendix}\label{sa}

We recall some tools we need on characteristic values of
finite meromorphic operator-valued functions. For more details on the subject, we refer 
for instance to \cite{go} and the book \cite[Section 4]{goh}. The content of this 
section follows \cite[Section 4]{goh}.

\medskip

Let $\mathscr{H}$ be separable Hilbert space, and let $\mathscr{L}(\mathscr{H})$ 
(resp. ${\rm GL}(\mathscr{H})$) denote the set of bounded (resp. invertible) linear 
operators in $\mathscr{H}$.

\begin{de}
Let $\mathcal{U}$ be a neighborhood of a fixed point $w \in \bc$, and 
$F : \mathcal{U} \setminus \lbrace w \rbrace \longrightarrow \mathscr{L}(\mathscr{H})$ 
be a holomorphic operator-valued function. The function $F$ is said to be finite 
meromorphic at $w$ if its Laurent expansion at $w$ has the form
\begin{equation}
F(z) = \sum_{n = m}^{+\infty} (z - w)^n A_n, \quad m > - \infty,
\end{equation}
where (if $m < 0$) the operators $A_m, \ldots, A_{-1}$ are of finite rank.
Moreover, if $A_0$ is a Fredholm operator, then the function $F$ is said to be Fredholm 
at $w$. In that case, the Fredholm index of $A_0$ is called the Fredholm index of $F$ 
at $w$.
\end{de}

\noindent
We have the following proposition:

\begin{prop}{\cite[Proposition 4.1.4]{goh}}\label{p,a1}
Let $\mathcal{D} \subseteq \mathbb{C}$ be a connected open set, $Z \subseteq \mathcal{D}$ 
be a closed and discrete subset of $\mathcal{D}$, and $F : \mathcal{D} \longrightarrow 
\mathscr{L}(\mathscr{H})$ be a holomorphic operator-valued function in $\mathcal{D} \backslash 
Z$. Assume that:
\begin{itemize}
\item $F$ is finite meromorphic on $\mathcal{D}$ (i.e. it is finite meromorphic near each 
point of $Z$),
\item $F$ is Fredholm at each point of $\mathcal{D}$,
\item there exists $w_0 \in \mathcal{D} \backslash Z$ such that $F(w_0)$ is invertible. 
\end{itemize}
Then, there exists a closed and discrete subset $Z'$ of $\mathcal{D}$ such that:
\begin{itemize}
\item $Z \subseteq Z'$,
\item $F(z)$ is invertible for each $z \in \mathcal{D} \backslash Z'$,
\item $F^{-1} : \mathcal{D} \backslash Z' \longrightarrow {\rm GL}(\mathscr{H})$ is finite
meromorphic and Fredholm at each point of $\mathcal{D}$.
\end{itemize}
\end{prop}

\noindent
In the setting of Proposition \ref{p,a1}, we define the characteristic values
of $F$ and their multiplicities as follows:

\begin{de}\label{d,a1}
The points of $Z'$ where the function $F$ or $F^{-1}$ is not holomorphic are called the
characteristic values of $F$. The multiplicity of a characteristic value $w_0$ is 
defined by
\begin{equation}
{\rm mult}(w_0) := \frac{1}{2i\pi} \textup{Tr} \oint_{\vert w - w_0 \vert = \rho} 
F'(z)F(z)^{-1} dz,
\end{equation}
where $\rho > 0$ is chosen small enough so that $\big\lbrace w \in \bc : \vert w - 
w_0 \vert \leq \rho \big\rbrace \cap Z' = \lbrace w_0 \rbrace$.
\end{de}

\noindent
According to Definition \ref{d,a1}, if the function $F$ is holomorphic in $\mathcal{D}$,
then the characteristic values of $F$ are just the complex numbers $w$ where the operator 
$F(w)$ is not invertible. Then, results of \cite{go} and \cite[Section 4]{goh} imply that 
${\rm mult}(w)$ is an integer.

\medskip

\noindent
Let $\Omega \subseteq \mathcal{D}$ be a connected domain with boundary $\partial \Omega$ 
not intersecting $Z'$. The sum of the multiplicities of the characteristic values of
the function $F$ lying in $\Omega$ is called {\it the index of $F$ with respect to the 
contour $\partial \Omega$} and is defined by 
\begin{equation}\label{eqa,2}
Ind_{\partial \Omega} \hspace{0.5mm} F := \frac{1}{2i\pi} \textup{Tr} 
\oint_{\partial \Omega} F'(z)F(z)^{-1} dz = \frac{1}{2i\pi} \textup{Tr} 
\oint_{\partial \Omega} F(z)^{-1} F'(z) dz.
\end{equation}

\medskip

\noindent
{\bf Acknowledgements:} O. Bourget is supported by the Chilean Fondecyt 
Grant $1161732$. D. Sambou is supported by the Chilean Fondecyt Grant $3170411$.

The authors express their gratitude to S. Golénia for bringing to their attention 
the paper \cite{af}, and to V. Bruneau and S. Kupin for their helpful discussions 
and valuable suggestions.


\end{document}